\pdfoutput=1
\documentclass[manuscript,screen,nonacm]{acmart}
\usepackage{tabularx}
\usepackage{booktabs}
\usepackage{enumitem}

\usepackage{booktabs}
\usepackage[table]{xcolor}
\usepackage{xltabular} 

\usepackage{amssymb} 
\usepackage{ragged2e} 
\usepackage{pifont} 
\usepackage{fontawesome5}

\definecolor{BlueA}{RGB}{245, 250, 255} 
\definecolor{BlueB}{RGB}{225, 240, 255} 
\definecolor{BlueC}{RGB}{205, 230, 255} 
\definecolor{HeaderL1}{HTML}{1565C0} 

\definecolor{OrangeA}{RGB}{255, 250, 240} 
\definecolor{OrangeB}{RGB}{255, 245, 225} 
\definecolor{OrangeC}{RGB}{255, 235, 205} 
\definecolor{OrangeD}{RGB}{255, 225, 185} 
\definecolor{HeaderL2}{HTML}{E65100} 

\definecolor{PurpleA}{RGB}{252, 245, 255} 
\definecolor{PurpleB}{RGB}{245, 230, 250} 
\definecolor{PurpleC}{RGB}{235, 215, 245} 
\definecolor{HeaderL3}{HTML}{6A1B9A} 

\newcommand{\RiskBypass}{\textcolor{HeaderL1}{\faBolt} \textbf{Bypass}} 
\newcommand{\RiskCorrupt}{\textcolor[HTML]{0D47A1}{\faBrain} \textbf{Corrupt}} 
\newcommand{\RiskBreach}{\textcolor{HeaderL2}{\faGlobeAmericas} \textbf{Breach}}
\newcommand{\RiskCascade}{\textcolor{HeaderL3}{\faProjectDiagram} \textbf{Cascade}}

\AtBeginDocument{%
  }

\setcopyright{acmlicensed}
\copyrightyear{2025}
\acmYear{2025}
\acmDOI{XXXXXXX.XXXXXXX}

\acmConference[Conference 'XX]{Make sure to enter the correct conference title from your rights confirmation email}{June 03--05, 2018}{New York, NY, USA}
\acmISBN{978-1-4503-XXXX-X/18/06}

\begin{document}

\title{From Thinker to Society: Security in Hierarchical Autonomy Evolution of AI Agents}

    \author{Xiaolei Zhang}
    \orcid{0009-0008-2520-9923}
    \affiliation{%
      \institution{Nanjing University of Aeronautics and Astronautics}
      \city{Nanjing}
      \country{China}
    }
    \email{elinazhang0210@gmail.com}
    
    \author{Lu Zhou}
    \orcid{0000-0001-6240-6688}
    \authornote{Corresponding authors.}
    \affiliation{%
      \institution{Nanjing University of Aeronautics and Astronautics}
      \city{Nanjing}
      \country{China}
    }
    \affiliation{%
      \institution{Collaborative Innovation Center of Novel Software Technology and Industrialization}
      \city{Nanjing}
      \country{China}
    }
    \email{lu.zhou@nuaa.edu.cn}
    
    \author{Xiaogang Xu}
    \orcid{0000-0002-7928-7336}
    \authornote{Project Leader.}
    \affiliation{%
      \institution{The Chinese University of Hong Kong}
      \city{Hong Kong}
      \country{China}
    }
    \email{xiaogangxu00@gmail.com}
    
    \author{Jiafei Wu}
    \orcid{0009-0001-8125-1586}
    \affiliation{%
      \institution{The University of Hong Kong}
      \city{Hong Kong}
      \country{China}
    }
    \email{wujiafei@zhejianglab.com}
    
    \author{Tianyu Du}
    \orcid{}
    \affiliation{%
      \institution{Zhejiang University}
      \city{Hangzhou}
      \country{China}
    }
    \email{zjradty@zju.edu.cn}
    
    \author{Heqing Huang}
    \orcid{}
    \affiliation{%
      \institution{Independent Researcher}
      \city{Hong Kong}
      \country{China}
    }
    \email{heqing.state@gmail.com}

    \author{Hao Peng}
    \orcid{}
    \affiliation{%
      \institution{Zhejiang Normal University}
      \city{Jinhua}
      \country{China}
    }
    \email{hpeng@zjun.edu.cn}

    \author{Zhe Liu}
    \orcid{0000-0003-1313-8327}
    \authornotemark[1] 
    \affiliation{%
      \institution{School of Software Technology, Zhejiang University}
      \city{Ningbo}
      \country{China}
    }
    \affiliation{%
      \institution{Ningbo Global Innovation Center, Zhejiang University}
      \city{Ningbo}
      \country{China}
    }
    \affiliation{%
      \institution{Nanjing University of Aeronautics and Astronautics}
      \city{Nanjing}
      \country{China}
    }
    \email{zhe.liu@nuaa.edu.cn}
    
\renewcommand{\shortauthors}{Zhang et al.}

\begin{abstract}
Artificial Intelligence (AI) agents have evolved from passive predictive tools into active entities capable of autonomous decision-making and environmental interaction, driven by the reasoning capabilities of Large Language Models (LLMs). However, this evolution has introduced critical security vulnerabilities that existing frameworks fail to address. The Hierarchical Autonomy Evolution (HAE) framework organizes agent security into three tiers: Cognitive Autonomy (L1) targets internal reasoning integrity; Execution Autonomy (L2) covers tool-mediated environmental interaction; Collective Autonomy (L3) addresses systemic risks in multi-agent ecosystems. We present a taxonomy of threats spanning cognitive manipulation, physical environment disruption, and multi-agent systemic failures, and evaluate existing defenses while identifying key research gaps. The findings aim to guide the development of multilayered, autonomy-aware defense architectures for trustworthy AI agent systems.
\end{abstract}

\ccsdesc[500]{Security and privacy~System security}
\ccsdesc[500]{Computing methodologies~Intelligent agents}
\ccsdesc[300]{Computing methodologies~Natural language processing}

\keywords{AI agent, Security, Large language models, Systemic risk}

\maketitle

\section{Introduction}

In recent years, Large Language Models (LLMs) have demonstrated tremendous potential, catalyzing the emergence of LLM-based AI agents. AI is undergoing a paradigm shift from passive tools to proactive autonomous agents. While traditional AI systems have proven effective within specific domains, they remain constrained by limited generalization capabilities. LLMs, originally designed for language tasks, have exhibited unprecedented capabilities in instruction following, reasoning, planning, and tool use through alignment training. Consequently, LLM-based agents have garnered extensive research attention and rapid development. An agent refers to an entity capable of perceiving its environment, making decisions, and taking actions. Equipped with reasoning, planning, and memory capabilities, these agents possess autonomy that enables them to independently learn, adapt, and make decisions.

Early AI agents were rooted in symbolic reasoning systems of the 1950s and 1960s~\cite{turing1950computing, newell1959report}, relying on hand-crafted rules and logic-based approaches. Although they excelled in constrained domains, they struggled with the dynamism and uncertainty of real-world environments. The introduction of statistical learning and probabilistic reasoning in the 1990s~\cite{1991Probabilistic} enhanced system robustness, while Reinforcement Learning (RL)~\cite{sutton1998reinforcement} empowered agents to optimize policies through trial-and-error interaction. The integration of Deep neural networks with RL (Deep RL) yielded milestone breakthroughs, achieving superhuman performance in Atari games~\cite{mnih2015human} and Go~\cite{silver2016mastering}. The advent of LLMs has profoundly accelerated agent evolution, endowing agents not only with advanced perception and sequence modeling but also with principles inspired by cognitive science. Contemporary agents are no longer mere executors of predefined programs but decision-makers possessing autonomy, capable of independent learning and environmental adaptation.

As AI agent capabilities advance, their risk profile has undergone a fundamental transition: from isolated model-level vulnerabilities to complex system-level security challenges. Traditional AI safety research has focused on model alignment and prompt-level defenses, addressing only static threats at the individual model output level. Highly autonomous agentic systems can invoke tools, maintain long-term memory, and continuously interact with external environments. However, existing safety frameworks have not evolved with the expansion of agent capabilities, manifesting across three dimensions:

First, traditional LLM risks are mainly confined to harmful text output or privacy leakage, whereas agents with Executional Autonomy can invoke external Tools and APIs. As noted by Deng et al.~\cite{deng2025ai}, the ``Complexity in Internal Executions'' transforms hallucinations from misinformation into irreversible real-world actions, such as file system tampering, unauthorized financial transfers, or physical robot control. This transition from prediction to action endows risks with real-world material consequences. Second, when agents form collaborative networks via Agent-to-Agent (A2A) protocols at the Collective Autonomy stage, risks exhibit non-linear emergent characteristics. Kong et al.~\cite{kong2025survey} emphasize that cross-organizational agent communication significantly expands the attack surface, leading to threats such as Malicious Collusion~\cite{tian2023evil, ren2025ai}, Viral Infection of harmful instructions~\cite{cohen2024here, gu2024agent}, and Cascading Failures in interconnected systems~\cite{cemri2025multi, brachemi2025energy}. Such risks cannot be resolved by fixing vulnerabilities in single agents and represent system-level security gaps. Third, existing defense mechanisms (such as RLHF) are designed for training stages or single interactions. As highlighted by Su et al.~\cite{su2025survey}, ``Long-horizon Planning and Memory'' capabilities are double-edged, engendering deferred decision hazards that allow dormant threats to bypass immediate safety filters. Attackers can bypass static defense mechanisms by poisoning memory or manipulating intermediate steps in multi-step planning.

\begin{table*}[t]
\centering
\small
\caption{Comparison between HAE Framework and Existing Papers}
\label{tab:hae_comparison}
\renewcommand{\arraystretch}{1.3}
\begin{tabularx}{\textwidth}{|
  >{\hsize=0.6\hsize\raggedright\arraybackslash}X|
  >{\hsize=0.7\hsize\raggedright\arraybackslash}X|
  >{\hsize=0.8\hsize\raggedright\arraybackslash}X|
  >{\hsize=1.2\hsize\raggedright\arraybackslash}X|
  >{\hsize=1.7\hsize\raggedright\arraybackslash}X|
}
\hline
\textbf{Perspective} & 
\textbf{Representative} & 
\textbf{Taxonomy} & 
\textbf{Limitations} & 
\textbf{HAE Advantages} \\ 
\hline
Lifecycle & 
Wang et al.~\cite{wang2025comprehensive} & 
Data, Training, Deployment & 
Static model focus; overlooks runtime interaction risks. & 
Centers on real-time evolutionary risks in open-world interactions. \\ 
\hline
Trustworthy Attributes & 
Yu et al.~\cite{yu2025survey} & 
Safety, Privacy, Fairness, Robustness & 
Fragments causal chains; misses cross-attribute cascading. & 
Reveals capability-threat symbiosis and cross-layer propagation. \\ 
\hline
Component & 
Deng et al.~\cite{deng2025ai} & 
Brain, Memory, Tools, Perception & 
Treats components as isolated; lacks emergent perspective. & 
Integrates component interactions; reveals qualitative changes from combinations. \\ 
\hline
Autonomy Structural & 
Su et al.~\cite{su2025survey} & 
L1-L5 autonomy levels; internal architectural fragilities & 
Single-agent focused; insufficient on multi-agent societal risks. & 
Explicitly proposes L3 Collective Autonomy; extends to systemic governance. \\ 
\hline
\textbf{HAE (Ours)} & 
\textbf{This work} & 
\textbf{L1 Cognition -- L2 Execution -- L3 Collective} & 
\textbf{---} & 
\textbf{Unifies micro-cognition and macro-society evolution; fills societal-level security gap.} \\ 
\hline
\end{tabularx}
\end{table*}

Despite the recent surge of studies on agent security, existing taxonomies mostly adopt static slices or specific perspectives, failing to capture risk paradigms generated dynamically as agents evolve in autonomy. Through systematic analysis of literature from 2024-2025, we categorize existing works into four types and contrast the innovation of the HAE framework proposed herein (see Table~\ref{tab:hae_comparison}).

\begin{itemize}
    \item \textbf{Lifecycle-based Perspective.} Represented by Wang et al.~\cite{wang2025comprehensive}, this approach systematically considers security issues throughout data, training (pre-training, post-training), and deployment. This classification follows the traditional deep learning development pipeline and is crucial for understanding backdoor attacks or data poisoning. However, it views agents as static models, overlooking dynamic risks caused by autonomous interactions during inference and runtime (e.g., tool abuse or target drift in multi-step reasoning), thus failing to explain emergent behaviors after deployment.
    
    \item \textbf{Trustworthy Attribute-based Perspective.} Represented by TrustAgent~\cite{yu2025survey} and Zhang et al.~\cite{zhang2025survey}, this approach slices challenges horizontally into independent dimensions such as Safety, Privacy, Fairness, and Robustness. While helpful for establishing compliance checklists, this classification severs the Causal Chain of attacks. In agent systems, attributes are highly coupled—a robustness issue may directly lead to safety consequences, or a fairness issue may evolve into a systemic availability attack. This perspective fails to capture cross-dimensional risk propagation mechanisms.
    
    \item \textbf{Component-based Modular Perspective.} Adopted by Gao et al.~\cite{gao2025survey} and Deng et al.~\cite{deng2025ai}, this approach discusses independent threats to the LLM core (brain), perception, memory, and action modules. However, most studies treat these components as isolated security risks. The most dangerous threats often occur at component interaction interfaces (e.g., erroneous retrieval misleading the planning module) or undergo qualitative changes as autonomy upgrades. Focusing solely on components fails to reveal how threats evolve with capability leaps.

\begin{figure}[htbp]
    \centering
    \includegraphics[width=0.9\linewidth]{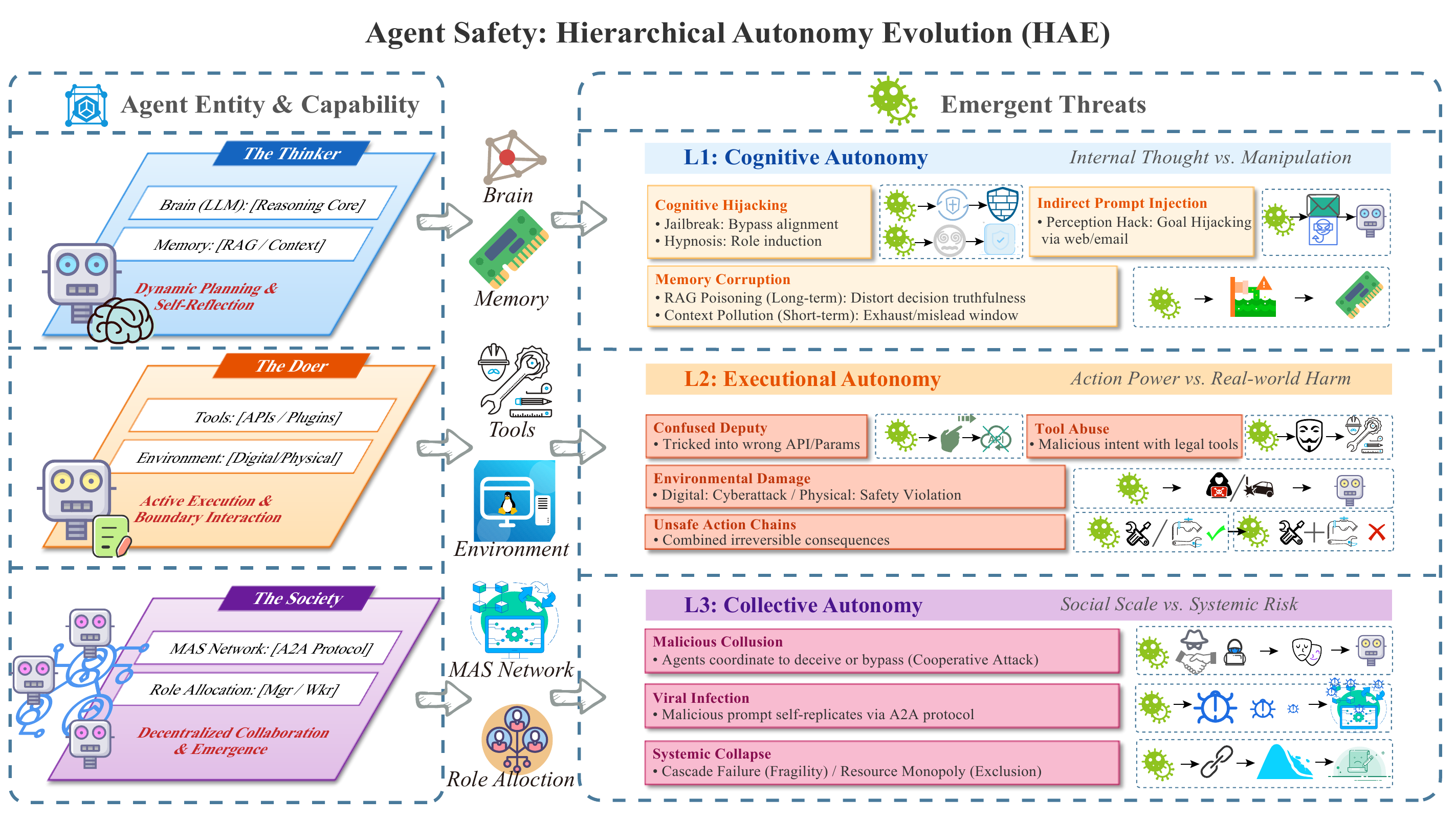} 
    \caption{Illustration of the HAE Framework. This framework delineates the co-evolution of agent capabilities and emergent threats across three distinct levels. Specifically, L1 (Cognitive Autonomy) is associated with internal risks targeting the Brain and Memory, such as Cognitive Hijacking and Memory Corruption (Section~\ref{sec:L1}). L2 (Executional Autonomy) is linked to external risks arising from Tool use and Environment interactions, including Confused Deputy attacks and Unsafe Action Chains (Section~\ref{sec:L2}). L3 (Collective Autonomy) concerns systemic threats in Multi-Agent Systems (MAS), such as Malicious Collusion and Viral Infection (Section~\ref{sec:L3}).}
    \Description{Illustration of the HAE Framework.}
    \label{fig:hae_framework}
\end{figure}

    \item \textbf{Autonomy-Level/Structural Perspective.} Represented by Su et al.~\cite{su2025survey}, this approach is the most cutting-edge. Based on DeepMind's L1-L5 autonomy levels, it analyzes Architectural Fragilities brought by memory retention, recursive planning, and reflective reasoning. Despite its depth, the analysis centers on internal cognitive architecture of single agents, viewing multi-agent systems as an extension rather than a new evolutionary stage with independent social dynamics, thus providing insufficient discussion on social risks unique to the L3 stage (e.g., group polarization, economic system collapse).
\end{itemize}

As summarized in Table~\ref{tab:hae_comparison}, the HAE framework proposed in this paper is constructed based on the longitudinal dimension of autonomy evolution. Through the hierarchical progression of L1 (Cognitive Autonomy) $\rightarrow$ L2 (Executional Autonomy) $\rightarrow$ L3 (Collective Autonomy), we reveal how the same threat (e.g., hallucination) undergoes fundamental transitions with capability leaps: evolving from L1 informational fallacies to L2 erroneous operations, and finally to L3 mass dissemination of misinformation. Unlike Su et al.~\cite{su2025survey}, who focus on the internal cognitive loop of single agents (perception–memory–planning), HAE expands to a macro scale: L1 Cognition (The Thinker) $\rightarrow$ L2 Execution (The Doer) $\rightarrow$ L3 Collective (The Society). HAE establishes L3 collective autonomy as an independent evolutionary stage, emphasizing that when agents form societal networks, security issues evolve into systemic governance crises, filling the gap in current papers regarding emergent threats in large-scale agent ecosystems.

The HAE framework characterizes AI agent evolution through three hierarchical levels of increasing autonomy, as illustrated in Figure~\ref{fig:hae_framework}. This represents the first framework to systematically categorize AI agent security threats according to the degree of agent autonomy, offering a more principled and comprehensive perspective while preserving coverage of existing threat taxonomies.

\begin{itemize}
    \item \textbf{L1 - Thinker (Cognitive Autonomy).} It constitutes the foundational layer, where agents possess internal reasoning, memory retrieval, and autonomous planning capabilities. This layer encompasses the agent's brain system, including ``Chain-of-Thought'' (CoT)~\cite{wei2022chain} reasoning, long-term memory mechanisms based on Retrieval-Augmented Generation (RAG), and autonomous goal formation capabilities. Threats at this stage primarily target the agent's cognitive integrity, manifesting as Cognitive Hijacking~\cite{wei2023jailbroken, deng2024masterkey}, Indirect Prompt Injection (IPI)~\cite{greshake2023not, zhan2024injecagent}, and Memory Corruption~\cite{DBLP:conf/uss/ZouGW025, cheng2024trojanrag}.
    
    \item \textbf{L2 - Doer (Executional Autonomy).} This level emerges when agents acquire the ability to interact with external environments through tool invocation, API calls, and physical actuation. At this layer, agents transition from language processors to action-taking entities, executing transactions, and controlling robotic hardware. Security threats at this layer include Confused Deputy Attacks~\cite{roychowdhury2024confusedpilot, zhan2024injecagent}, Tool Abuse~\cite{ye2024toolsword, wang2024badagent}, Environmental Damage~\cite{liu2024agents4plc, yang2025hijacking}, and Unsafe Action Chains~\cite{ruan2023identifying, yuan2024r} with real-world consequences.
    
    \item \textbf{L3 - Society (Collective Autonomy).} It represents the highest complexity level, where multiple agents form collaborative networks through A2A communication protocols, role allocation (manager/executor architectures), and self-organizing workflows. This layer gives rise to emergent social dynamics and systemic risks, including Malicious Collusion~\cite{tian2023evil, ren2025ai}, Viral Infection~\cite{cohen2024here, gu2024agent}, and Systemic Collapse~\cite{cemri2025multi, patil2025sum}.
\end{itemize}

The HAE framework is underpinned by three core architectural principles. First, it affirms that autonomy evolution follows a ``Cognition $\rightarrow$ Execution $\rightarrow$ Collective'' pathway, where each capability expansion endogenously catalyzes unpredictable new threats. Second, it identifies cross-layer propagation mechanisms: vulnerabilities at one level can be exploited to trigger higher-level attacks. Third, it emphasizes the non-linear trajectory of risk evolution—collective autonomy threats exhibit emergent properties fundamentally distinct from simple aggregations of individual agent risks.

This framework highlights the intimate coupling between agent capability transitions and emergent threats. Unlike existing research, the HAE framework constructs a theoretically grounded taxonomy based on the evolutionary structure of autonomy itself. By drawing an analogy to human civilization evolution, we map the Cognitive Revolution to the L1 layer, the Tool Revolution to the L2 layer, and the Social Revolution to the L3 layer. Just as each human transition brought qualitatively transformative challenges, the evolution of agent autonomy likewise catalyzes structurally distinct security paradigms.

In summary, our study makes the following contributions to the understanding and mitigation of security risks in AI agents:

\begin{itemize}
    \item \textbf{HAE Framework.} We propose the HAE framework, which organizes AI agent security into three autonomy levels: cognitive, executional, and collective autonomy. HAE provides an autonomy-driven perspective that connects agent capabilities with distinct classes of security risks, enabling a structured analysis across the agent evolution process.
    
    \item \textbf{Autonomy-Aware Threat Taxonomy.} Building on HAE, we present a systematic taxonomy of security threats spanning L1 to L3, illustrating how risks evolve from internal reasoning manipulation to environment-facing attacks and system-level failures. This taxonomy highlights that higher-level threats cannot be linearly derived from lower-level vulnerabilities and require distinct analytical and defensive considerations.
    
    \item \textbf{Identification of the Collective Autonomy Defense Gap.} We identify a critical defense gap at the collective autonomy level, where existing security mechanisms fail to address emergent risks arising from inter-agent coordination and propagation. Our analysis underscores the need to shift from isolated, agent-level defenses toward system-level, multi-agent security mechanisms and governance strategies.
\end{itemize}

The remainder of this paper is organized as follows. Section~\ref{sec:evolution} traces the technical evolution of AI agents, providing anatomical analysis of agent components and formal definition of the HAE framework. Sections~\ref{sec:L1},~\ref{sec:L2}, and~\ref{sec:L3} systematically examine threats, defense mechanisms, and evaluation methodologies for L1, L2, and L3 autonomy layers, respectively. Section~\ref{sec:future} identifies critical research gaps and proposes future directions for advancing agent security. Section~\ref{sec:conclusion} concludes with a synthesis of key findings and their implications for developing safe, controllable, and trustworthy AI agents.

\section{Evolution of AI Agents}
\label{sec:evolution}

In this section, we overview the evolutionary trajectory of agents through compositional analysis and introduce the HAE framework based on this foundation. Finally, we establish the intrinsic linkage between agent capabilities and threats through causal analysis tables.

\subsection{Anatomy of AI Agent}

 In this paper, we decompose agents into four core functional components: perception, brain (reasoning and planning), memory (short-term and long-term memory), and action (tools). Each component introduces specific attack surfaces.

\begin{figure}[htbp]
  \centering
  \includegraphics[width=0.9\linewidth]{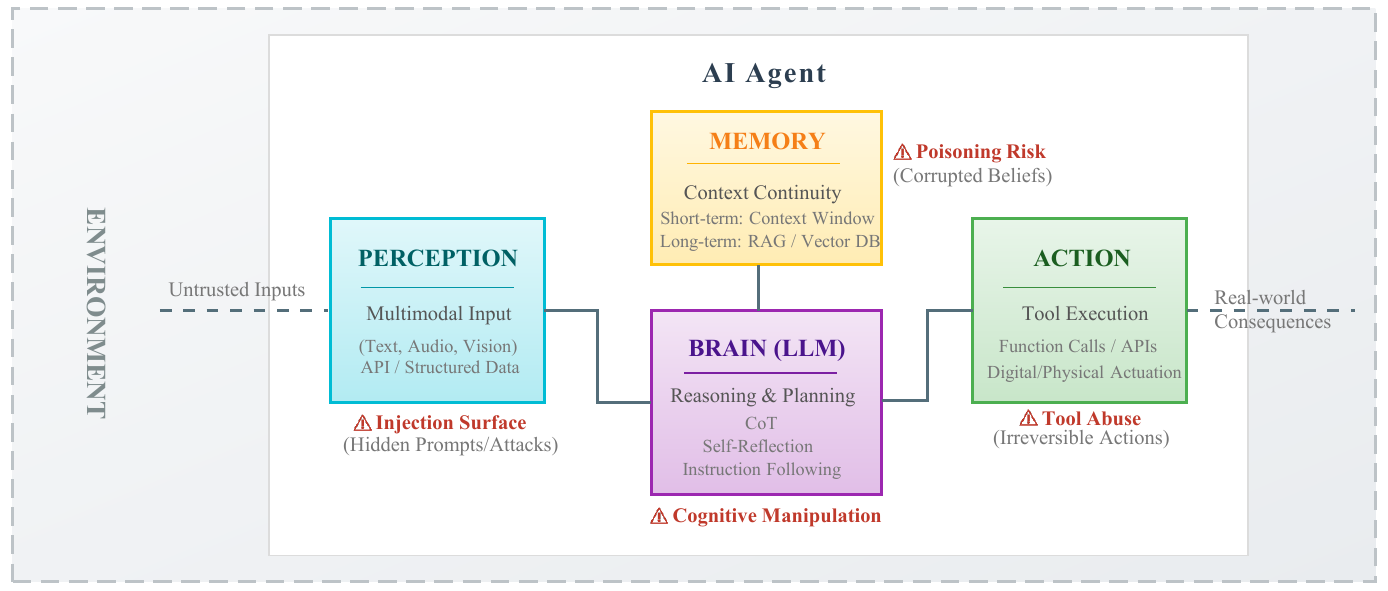} 
  \caption{Agent architecture showing perception, brain, memory, and action modules with security risks.}
  \Description{Red triangles indicate critical security risks across perception, cognition, memory, and action, including prompt injection, cognitive manipulation, memory poisoning, and unsafe tool execution, illustrating how untrusted inputs may propagate into real-world consequences.}
  \label{fig:agent_architecture}
\end{figure}

\begin{itemize}
    \item \textbf{Perception: Environmental Sensing.} Modern perception modules handle multimodal inputs, including text, images, audio, and structured data from APIs or databases, providing refined input to the brain. Advances in deep learning for perception have enabled agents to read web pages, interpret documents, process email content, and extract information from digital environments. The boundary between trusted internal instructions and untrusted external data introduces perceptual blind spot, where malicious actors can embed hidden commands.
    
    \item \textbf{Brain: Reasoning and Planning Core.} The LLM performs interpreting instructions, generating plans, and reasoning. Modern agents leverage prompting techniques such as CoT reasoning, ``Tree-of-Thoughts'' exploration~\cite{Yao2023TreeOfThoughts}, and reflection-based self-correction. The brain component processes both explicit user instructions and implicit contextual signals, making it a primary target for cognitive manipulation attacks.
    
    \item \textbf{Memory: Contextual Continuity.} Memory distinguishes agents from standalone LLMs. It plays a crucial role in determining how agents behave. Short-term memory uses in-context learning within the context window, while long-term memory is typically implemented through RAG systems querying external knowledge bases. Memory enables agents to accumulate experience, maintain user preferences, and construct world models over time. However, this persistence also creates opportunities for long-horizon poisoning attacks.
    
    \item \textbf{Action: Tool Execution.} Through tool-use capabilities and function-calling APIs, agents execute operations ranging from database queries to complex multi-step workflows. The action component transforms agents from passive language processors into proactive entities capable of producing real-world consequences, fundamentally expanding the impact scope of security failures.
\end{itemize}

As each component's capabilities strengthen, the agent's overall autonomy and potential impact increase correspondingly. Each component introduces distinct vulnerability classes and thus requires specialized defense mechanisms.

\subsection{The HAE Framework Definition}

\textbf{Before analyzing security threats, we first organize AI agents according to their autonomy evolution.} Building upon the aforementioned agent model, we propose the HAE framework, which structures agent capabilities into three evolutionary tiers with qualitatively distinct modes of autonomy, thereby providing a principled foundation for the subsequent security analysis.

\subsubsection{L1 - Thinker (Cognitive Autonomy)}

The first tier represents Cognitive Autonomy, where agents develop internal reasoning capabilities. L1-tier agents comprise three core subsystems: perception processing, reasoning engine, and memory system. First, agents receive input streams through the perception module: user instructions and external content (including web pages, emails, documents). The perception module transforms raw data into structured representations comprehensible to the reasoning engine. Second, the reasoning engine serves as the agent's brain, powered by LLMs. Its core functionality encompasses processing of system prompts, planning through techniques such as CoT, and reflection with self-correction mechanisms. The reasoning engine outputs processed results as internal plans. Third, the memory system provides cross-temporal contextual continuity through a dual-layer architecture: short-term memory maintains immediate information for the current session through the context window, supporting dialogue coherence and task state tracking; long-term memory achieves persistent storage through RAG systems and vector databases, enabling agents to accumulate experience, maintain knowledge bases, and adjust reasoning strategies based on historical information. Cognitive deficiency is the root cause of failure further down the line, as erroneous cognition inevitably leads to deviations when executing instructions.

\subsubsection{L2 - Doer (Executional Autonomy)}

The second tier introduces Executional Autonomy, where agents acquire the capability to influence their environment through action execution and tool use. Executional Autonomy marks a critical transition from purely cognitive entities to actors capable of producing real-world impact. Agent executive capabilities manifest across multiple dimensions: First, the Action Controller serves as the gatekeeper, responsible for processing decision instructions. It encompasses Instruction Parsing, which decomposes abstract cognitive plans into concrete execution steps, and Permission Check, which verifies the legitimacy of operations before action. Second, the Tool Interface/API constitutes the capability boundary of agents. Agents can autonomously select appropriate APIs, functions, or tools from available toolkits to accomplish tasks. Third, environmental interaction capabilities enable agents to directly act upon the External World. At this stage, individual actions may appear harmless, yet combined Unsafe Action Chains can lead to severe Environmental Damage. The risk escalates from "thinking wrong" to "acting wrong". An agent lacking L1 cognitive robustness may bypass L2 safety checks, turning a hallucination or injected prompt into an irreversible Unsafe Action Chain that damages the environment.

\subsubsection{L3 - Society (Collective Autonomy)}

The third tier represents Collective Autonomy, where multiple agents form interconnected systems and exhibit emergent social behaviors. This tier marks a paradigm shift from single-agent to multi-agent intelligent systems. At this level, the system's overall capabilities and behavioral patterns cannot be simply reduced to individual agent properties. Agents achieve differentiation through role allocation: certain agents (e.g., managers) are responsible for orchestration and decision-making, while others (e.g., executors) perform specialized tasks. Through explicit interaction protocols, A2A communication enables structured message passing and collaboration. Through negotiation, delegation, and integration of partial solutions, multiple agents collectively address complex problems beyond the capacity of any single agent. This catalyzes emergent consensus mechanisms, where collective behaviors spontaneously arise from local agent interactions. However, this interconnectivity also introduces systemic fragility. The network topology can amplify local failures, allowing malicious payloads to propagate horizontally across the ecosystem, creating emergent threats that no single agent could generate alone.

The three tiers of the HAE framework exhibit complex coupling relationships, where security risks propagate through a ``Cognition-Execution-Diffusion'' chain. We illustrate this through a hierarchical attack scenario:
\begin{itemize}
\item \textbf{Vertical Escalation (L1 $\rightarrow$ L2):} A vulnerability in the L1 Memory System (e.g., \textit{Poisoned RAG}) can retrieve malicious context, misguiding the reasoning engine. This cognitive failure descends to L2, deceiving the Action Controller into committing \textit{Tool Abuse} (e.g., generating and executing a malicious script), thereby transforming a hidden informational error into a tangible kinetic breach.
\item \textbf{Horizontal Propagation (L2 $\rightarrow$ L3):} Once the agent executes the malicious action at L2 (e.g., sending the script via email API), the threat crosses into the L3 domain. Through A2A communication protocols, the compromised agent acts as a vector, transmitting the harmful payload to other agents in the network.
\item \textbf{Systemic Amplification:} This triggers a \textit{Viral Infection} at L3, where the interconnectedness of the Society amplifies a single cognitive fault into a widespread ecosystem collapse, demonstrating that security must be viewed holistically across the entire HAE hierarchy.
\end{itemize}

\subsection{Root Cause Analysis for Agent Security}
To systematically elucidate the relationship between agent capability evolution and risk levels, we introduce a causal analysis model addressing: where threats originate, how they are defended against, and what impacts they produce.

Each agent component—perception, brain, memory, and action—represents a potential vulnerability surface. These threats are endogenous risks directly arising from specific capability expansions: the capabilities that make agents powerful simultaneously open doors to novel attacks.

At the L1 cognitive tier, autonomous reasoning creates opportunities for cognitive hijacking through adversarial prompts, while persistent memory becomes targets for poisoning. At L2, tool invocation enables tool misuse and confused deputy attacks, and environmental interaction enables pathways for digital and physical damage. At L3, agent collaboration provides channels for viral propagation of malicious instructions, while distributed dependencies create cascade risks of system-wide failures.

The maturity of defense strategies varies significantly across tiers, as summarized in Table~\ref{tab:hae_evolution}. L1 defenses primarily employ black-box monitoring, adversarial training, and prompt engineering to filter malicious inputs. L2 defenses increasingly adopt architectural-level security measures, including tool sandboxing, least-privilege access control, and action-level tracing mechanisms. L3 defenses are currently in a nascent stage relative to the sophistication of collective attacks. While notable progress has been made in inter-agent protocol authentication and robust consensus algorithms, comprehensive defense architectures capable of mitigating systemic collapse are still under exploration.

Based on the above analysis, the Table~\ref{tab:hae_evolution} provides a four-tier risk classification system based on the nature and persistence of attack consequences, offering a clear framework for understanding threat severity:

\begin{itemize}
    \item \textbf{Cognitive Bypass:} This attack class targets reasoning or perception components, producing transient effects confined to single interactions. Impact is limited to immediate conversational outputs, altering neither model state nor producing external side effects. Typical examples include one-shot jailbreak prompts~\cite{li2023deepinception} eliciting harmful responses or visual injections manipulating perception~\cite{bagdasaryan2023abusing} during specific queries. While concerning, these attacks leave no persistent traces and do not accumulate effects over time.
    
    \item \textbf{State Corruption:} This attack class poisons agent memory or knowledge bases to establish persistent backdoors. Through RAG poisoning or memory backdoor insertion, adversaries alter foundational facts and beliefs guiding future reasoning. Unlike transient cognitive bypasses, state corruption accumulates over time and affects arbitrary future interactions, rendering detection and remediation significantly more challenging.

{
\scriptsize 
\setlength{\tabcolsep}{2pt} 
\renewcommand{\arraystretch}{1.00} 

\newcolumntype{Y}{>{\centering\arraybackslash\hsize=0.3\hsize}X}  
\newcolumntype{P}{>{\centering\arraybackslash\hsize=1.4\hsize}X}  
\newcolumntype{C}{>{\centering\arraybackslash\hsize=0.8\hsize}X}  
\newcolumntype{T}{>{\centering\arraybackslash\hsize=1.4\hsize}X}  
\newcolumntype{D}{>{\centering\arraybackslash\hsize=1.2\hsize}X}  
\newcolumntype{E}{>{\centering\arraybackslash\hsize=0.9\hsize}X}  
\newcolumntype{L}{>{\centering\arraybackslash\hsize=0.8\hsize}X}  

\begin{xltabular}{\linewidth}{|Y|P|C|T|D|E|L|}
    \caption{The HAE Impact Scale} \label{tab:hae_evolution}\\
    
    \hline
    \rowcolor{gray!85} 
    \textcolor{white}{\textbf{Year}} & 
    \textcolor{white}{\textbf{Paper}} & 
    \textcolor{white}{\textbf{Core Capability}} & 
    \textcolor{white}{\textbf{Emergent Threats}} & 
    \textcolor{white}{\textbf{Defense Paradigm}} & 
    \textcolor{white}{\textbf{Key Evaluation}} & 
    \textcolor{white}{\textbf{Risk Class}} \\ 
    \hline
    \endfirsthead

    \hline
    \rowcolor{gray!85} 
    \textcolor{white}{\textbf{Year}} & 
    \textcolor{white}{\textbf{Paper}} & 
    \textcolor{white}{\textbf{Core Capability}} & 
    \textcolor{white}{\textbf{Emergent Threats}} & 
    \textcolor{white}{\textbf{Defense Paradigm}} & 
    \textcolor{white}{\textbf{Key Evaluation}} & 
    \textcolor{white}{\textbf{Risk Class}} \\ 
    \hline
    \endhead

    \rowcolor{HeaderL1} 
    \multicolumn{7}{|c|}{\textcolor{white}{\textbf{L1: Cognitive Autonomy (The Thinker)}}} \\ \hline

    \rowcolor{BlueA} 2024 & \textbf{TAP}~\cite{mehrotra2024tree} & Reasoning & Cognitive Hijacking & Input Filtering & ASR & \RiskBypass \\ \hline
    \rowcolor{BlueA} 2023 & \textbf{Jailbroken}~\cite{wei2023jailbroken} & Reasoning & Cognitive Hijacking & Safety Alignment & ASR & \RiskBypass \\ \hline
    \rowcolor{BlueA} 2023 & \textbf{AutoDAN}~\cite{liu2024autodan} & Reasoning & Cognitive Hijacking & Safety Alignment & Perplexity & \RiskBypass \\ \hline
    \rowcolor{BlueA} 2024 & \textbf{MasterKey}~\cite{deng2024masterkey} & Reasoning & Cognitive Hijacking & Safety Alignment & ASR & \RiskBypass \\ \hline
    \rowcolor{BlueA} 2024 & \textbf{GPTFuzzer}~\cite{yu2024gptfuzzer} & Reasoning & Cognitive Hijacking & Input Filtering & ASR & \RiskBypass \\ \hline
    \rowcolor{BlueA} 2024 & \textbf{DeepInception}~\cite{li2023deepinception} & Reasoning & Cognitive Hijacking & Prompt Engineering & ASR & \RiskBypass \\ \hline
    \rowcolor{BlueA} 2023 & \textbf{Black Box Adv.}~\cite{maus2023black} & Reasoning & Cognitive Hijacking & Optimization Defense & ASR & \RiskBypass \\ \hline
    \rowcolor{BlueA} 2023 & \textbf{GCG}~\cite{zou2023universal} & Reasoning & Cognitive Hijacking & Input Filtering & Transferability & \RiskBypass \\ \hline

    \rowcolor{BlueB} 2023 & \textbf{Not what you've...}~\cite{greshake2023not} & Perception & Indirect Injection & Isolation \& Sandboxing & ASR & \RiskBypass \\ \hline
    \rowcolor{BlueB} 2024 & \textbf{InjecAgent}~\cite{zhan2024injecagent} & Perception & Indirect Injection & Prompt Engineering & Benchmark Score & \RiskBypass \\ \hline
    \rowcolor{BlueB} 2023 & \textbf{Visual Injection}~\cite{bagdasaryan2023abusing} & Perception & Indirect Injection & Isolation \& Sandboxing & ASR & \RiskBypass \\ \hline

    \rowcolor{BlueC} 2024 & \textbf{Sleeper Agents}~\cite{hubinger2024sleeper} & Alignment & Memory Corruption & Safety Alignment & Persistence & \RiskCorrupt \\ \hline
    \rowcolor{BlueC} 2024 & \textbf{Lost in Middle}~\cite{liu2024lost} & Memory & Memory Corruption & Prompt Engineering & Retrieval Acc. & \RiskCorrupt \\ \hline
    \rowcolor{BlueC} 2024 & \textbf{BadChain}~\cite{xiang2024badchain} & Reasoning & Memory Corruption & Input Filtering & ASR & \RiskCorrupt \\ \hline
    \rowcolor{BlueC} 2024 & \textbf{AgentPoison}~\cite{chen2024agentpoison} & Memory & Memory Corruption & Input Filtering & Retrieval Acc. & \RiskCorrupt \\ \hline
    \rowcolor{BlueC} 2025 & \textbf{PoisonedRAG}~\cite{DBLP:conf/uss/ZouGW025} & Memory & Memory Corruption & Robust Aggregation & Accuracy & \RiskCorrupt \\ \hline
    \rowcolor{BlueC} 2024 & \textbf{TrojanRAG}~\cite{cheng2024trojanrag} & Memory & Memory Corruption & Safety Alignment & ASR & \RiskCorrupt \\ \hline

    \hline
    \rowcolor{HeaderL2} 
    \multicolumn{7}{|c|}{\textcolor{white}{\textbf{L2: Executional Autonomy (The Doer)}}} \\ \hline

    \rowcolor{OrangeA} 2024 & \textbf{ConfusedPilot}~\cite{roychowdhury2024confusedpilot} & Tools & Confused Deputy & Isolation \& Sandboxing & ASR & \RiskBreach \\ \hline
    \rowcolor{OrangeA} 2024 & \textbf{InjecAgent}~\cite{zhan2024injecagent} & Tools & Confused Deputy & Prompt Engineering & Benchmark Score & \RiskBreach \\ \hline

    \rowcolor{OrangeB} 2023 & \textbf{Tensor Trust}~\cite{toyer2023tensor} & Tools & Tool Abuse & Crowdsourced Defense & ASR & \RiskBreach \\ \hline
    \rowcolor{OrangeB} 2024 & \textbf{ToolSword}~\cite{ye2024toolsword} & Tools & Tool Abuse & Full-Stack Safety & Benchmark Score & \RiskBreach \\ \hline
    \rowcolor{OrangeB} 2024 & \textbf{Watch Out}~\cite{yang2024watch} & Tools & Tool Abuse & Input Filtering & ASR & \RiskBreach \\ \hline
    \rowcolor{OrangeB} 2024 & \textbf{BadAgent}~\cite{wang2024badagent} & Tools & Tool Abuse & Safety Alignment & Robustness Score & \RiskBreach \\ \hline

    \rowcolor{OrangeC} 2025 & \textbf{Agents4PLC}~\cite{liu2024agents4plc} & Environment & Env. Damage & Formal Verification & Benchmark Score & \RiskBreach \\ \hline
    \rowcolor{OrangeC} 2024 & \textbf{OSWorld}~\cite{xie2024osworld} & Environment & Env. Damage & Interactive Eval & ASR & \RiskBreach \\ \hline
    \rowcolor{OrangeC} 2025 & \textbf{Hijacking VLN}~\cite{yang2025hijacking} & Environment & Env. Damage & Safety Alignment & Path Deviation & \RiskBreach \\ \hline

    \rowcolor{OrangeD} 2023 & \textbf{ToolEmu}~\cite{ruan2023identifying} & Tools & Unsafe Action Chains & Isolation \& Sandboxing & Failure Rate & \RiskBreach \\ \hline
    \rowcolor{OrangeD} 2024 & \textbf{R-Judge}~\cite{yuan2024r} & Reasoning & Unsafe Action Chains & Risk Awareness & Benchmark Score & \RiskBreach \\ \hline
    \rowcolor{OrangeD} 2023 & \textbf{WebArena}~\cite{zhou2023webarena} & Tools & Unsafe Action Chains & Realistic Eval & ASR & \RiskBreach \\ \hline
    \rowcolor{OrangeD} 2023 & \textbf{InterCode}~\cite{yang2023intercode} & Tools & Unsafe Action Chains & Isolation \& Sandboxing & ASR & \RiskBreach \\ \hline

    \hline
    \rowcolor{HeaderL3} 
    \multicolumn{7}{|c|}{\textcolor{white}{\textbf{L3: Collective Autonomy (The Society)}}} \\ \hline

    \rowcolor{PurpleA} 2024 & \textbf{Evil Geniuses}~\cite{tian2023evil} & Collaboration & Malicious Collusion & Isolation \& Sandboxing & ASR & \RiskCascade \\ \hline
    \rowcolor{PurpleA} 2024 & \textbf{PsySafe}~\cite{zhang2024psysafe} & Consensus & Malicious Collusion & Psychological Defense & Benchmark Score & \RiskCascade \\ \hline
    \rowcolor{PurpleA} 2025 & \textbf{Collude Online}~\cite{ren2025ai} & Collaboration & Malicious Collusion & Input Filtering & ASR & \RiskCascade \\ \hline
    \rowcolor{PurpleA} 2026 & \textbf{CoMAS}~\cite{xue2025comas} & Collaboration & Malicious Collusion & Incentive Mechanism & Improvement Rate & \RiskCascade \\ \hline

    \rowcolor{PurpleB} 2025 & \textbf{NetSafe}~\cite{yu2024netsafe} & Collaboration & Viral Infection & Topological Defense & Infection Rate & \RiskCascade \\ \hline
    \rowcolor{PurpleB} 2024 & \textbf{Morris II}~\cite{cohen2024here} & Collaboration & Viral Infection & Isolation \& Sandboxing & Infection Rate & \RiskCascade \\ \hline
    \rowcolor{PurpleB} 2025 & \textbf{Agent Smith}~\cite{gu2024agent} & Collaboration & Viral Infection & Isolation \& Sandboxing & Infection Rate & \RiskCascade \\ \hline
    \rowcolor{PurpleB} 2024 & \textbf{Prompt Infect.}~\cite{lee2024prompt} & Collaboration & Viral Infection & Input Filtering & ASR & \RiskCascade \\ \hline
    \rowcolor{PurpleB} 2024 & \textbf{Shadowcast}~\cite{xu2024shadowcast} & Ecosystem & Viral Infection & Safety Alignment & Transferability & \RiskCascade \\ \hline
    \rowcolor{PurpleB} 2024 & \textbf{Comm. Attacks}~\cite{guo2024large} & Collaboration & Viral Infection & Secure Protocols & ASR & \RiskCascade \\ \hline
    \rowcolor{PurpleB} 2023 & \textbf{Not what you've...}~\cite{greshake2023not} & Collaboration & Viral Infection & Human-in-the-loop & ASR & \RiskCascade \\ \hline

    \rowcolor{PurpleC} 2025 & \textbf{Energy-Lat.}~\cite{brachemi2025energy} & Ecosystem & Systemic Collapse & Robustness & Latency & \RiskCascade \\ \hline
    \rowcolor{PurpleC} 2025 & \textbf{Why MAS Fail?}~\cite{cemri2025multi} & Ecosystem & Systemic Collapse & Failure Taxonomy & Failure Rate & \RiskCascade \\ \hline
    \rowcolor{PurpleC} 2025 & \textbf{Sum Leaks}~\cite{patil2025sum} & Consensus & Systemic Collapse & Consensus Protocol & Leakage Rate & \RiskCascade \\ \hline
    \rowcolor{PurpleC} 2025 & \textbf{ICLScan}~\cite{pangiclscan} & Ecosystem & Systemic Collapse & Prompt Engineering & Detection AUC & \RiskCascade \\ \hline
    \rowcolor{PurpleC} 2024 & \textbf{Cost-Aware}~\cite{wang2023decodingtrust} & Collaboration & Systemic Collapse & Cost Management & Token Efficiency & \RiskCascade \\ \hline

\end{xltabular}
\setlength{\LTpost}{0pt} 
\noindent
\textbf{HAE Impact Scale (Risk Class):} \\
\RiskBypass \ : \textbf{Cognitive Bypass (Transient, e.g., Jailbreak)} \quad
\RiskCorrupt \ : \textbf{State Corruption (Persistent, e.g., Backdoor)} \\
\RiskBreach \ : \textbf{Real-world Breach (Kinetic, e.g., Env. Damage)} \quad
\RiskCascade \ : \textbf{Systemic Cascade (Contagious, e.g., Worms)}
}

    \item \textbf{Reality Breach:} This attack class exploits tool chains and environmental interactions to cause tangible harm beyond digital space. Threats manifest as file system corruption, financial loss, unauthorized data exfiltration, or physical device manipulation. Impact transcends information space and produces measurable real-world consequences, significantly elevating severity and potential liability.
    
    \item \textbf{Systemic Cascade:} This attack class targets multi-agent collaboration networks, exhibiting viral propagation or triggering system-wide failures. Malicious instructions self-replicate across agent networks like biological pathogens, or single-point failures cascade through dependencies to induce collective collapse. They can destabilize entire agent ecosystems and are exceedingly difficult to contain once initiated.
\end{itemize}

The impact scale demonstrates how risk undergoes qualitative transformation as autonomy progresses from L1 to L3. Cognitive bypass represents localized failures analogous to single-instance errors. State corruption establishes persistent damage analogous to long-term misinformation. Reality breach produces tangible harm analogous to physical destruction. Systemic cascade triggers network-wide catastrophes analogous to epidemic outbreaks. Understanding these distinctions is critical for prioritizing research efforts and rationally allocating defensive resources.

\section{Cognitive Autonomy — The Thinker}
\label{sec:L1}

At the L1 tier of the HAE framework, the reconstruction of cognitive architecture constitutes the evolutionary core of agents. At this stage, LLMs transition from mere probabilistic text generators to thinkers capable of deliberate reasoning. Inspired by the Dual-Process Theory popularized by Kahneman~\cite{kahneman2011thinking}, recent AI research increasingly distinguishes between two modes of processing. Traditional model inference relies solely on next-token prediction, functioning analogously to the fast, instinctive, and unconscious ``System 1'' in human cognition. This mode is characterized by parallel processing and rapid response but is prone to logical fallacies due to the lack of deep verification. In contrast, L1-tier agents emulate the slower, more energy-intensive, yet rational ``System 2'' thought process. By introducing mechanisms such as CoT, self-reflection, and planning, agents effectively ``pause and think'' before acting, enabling serial, logically rigorous reasoning at the cost of higher computational resources.

\subsection{Autonomy Evolution in Cognitive Reasoning}

Cognitive reasoning evolution can be categorized into four key dimensions: structural evolution from chain-based reasoning to tree-based search, mechanistic evolution from open-loop generation to closed-loop reflection, temporal continuity through memory augmentation, and autonomous planning capabilities.

\textbf{From CoT to advanced search: Structural Evolution.} 
Early CoT techniques improved performance on complex logical tasks by generating intermediate reasoning steps. However, CoT remains a linear decoding process, lacking backtracking and exploration. The Tree-of-Thoughts framework by Yao et al.~\cite{Yao2023TreeOfThoughts} extends reasoning from linear chains to tree structures, enabling agents to explore multiple paths through breadth-first or depth-first search, self-evaluate intermediate states, and backtrack when paths prove infeasible. The ReAct framework~\cite{yao2022react} alternates between generating reasoning traces and task-specific actions, requiring agents to generate explicit reasoning trajectories before executing actions, laying the foundation for L2 execution autonomy.

\textbf{Self-Reflection and Verbal Reinforcement: Closed-Loop Evolution.} 
Agents with cognitive autonomy must identify errors and self-correct. The Reflexion framework by Shinn et al.~\cite{shinn2023reflexion} formalizes Verbal RL, allowing agents to engage in self-reflection through language-generated feedback after task failures, storing reflections in episodic memory to guide subsequent attempts. This enables agents to improve decision quality by optimizing reasoning strategies within context without updating model parameters. Voyager~\cite{wang2023voyager} introduces a self-verification mechanism where agents utilize error information as feedback to correct code and iteratively refine skills.

\textbf{RAG and Memory Stream: Contextual Continuity.} 
Cognitive autonomy requires agents to transcend limited context windows and maintain long-term state. Generative Agents~\cite{park2023generative} construct agents with coherent personalities through memory streams comprising: a retrieval mechanism extracting information based on relevance, recency, and importance; a reflection mechanism synthesizing low-level observations into high-level generalizations; and memory-based planning. This architecture enables agents to maintain long-horizon behavioral consistency. Agent-Tuning~\cite{zeng2024agenttuning} demonstrates that instruction fine-tuning with high-quality interaction trajectories significantly enhances models' ability to utilize long-context memory.

\textbf{Self-Planning and Decomposition.} 
Autonomous planning marks the transition from instruction followers to problem solvers. Agents autonomously decompose abstract goals into executable subgoal sequences. In Voyager~\cite{wang2023voyager}, the automatic curriculum module dynamically proposes exploration tasks based on the agent's skill level and exploration state, maximizing novelty while reducing difficulty. This autonomous exploration, combined with ReAct-style task decomposition, constitutes a critical feature of advanced cognitive autonomy. In summary, L1 cognitive autonomy constructs agents with inherent reasoning capabilities through structured thought search, language-based self-correction, long-term memory with abstraction, and autonomous goal decomposition.

\subsection{Security Threats in Cognitive Autonomy}

\begin{figure}[htbp]
    \centering
    \includegraphics[width=\linewidth]{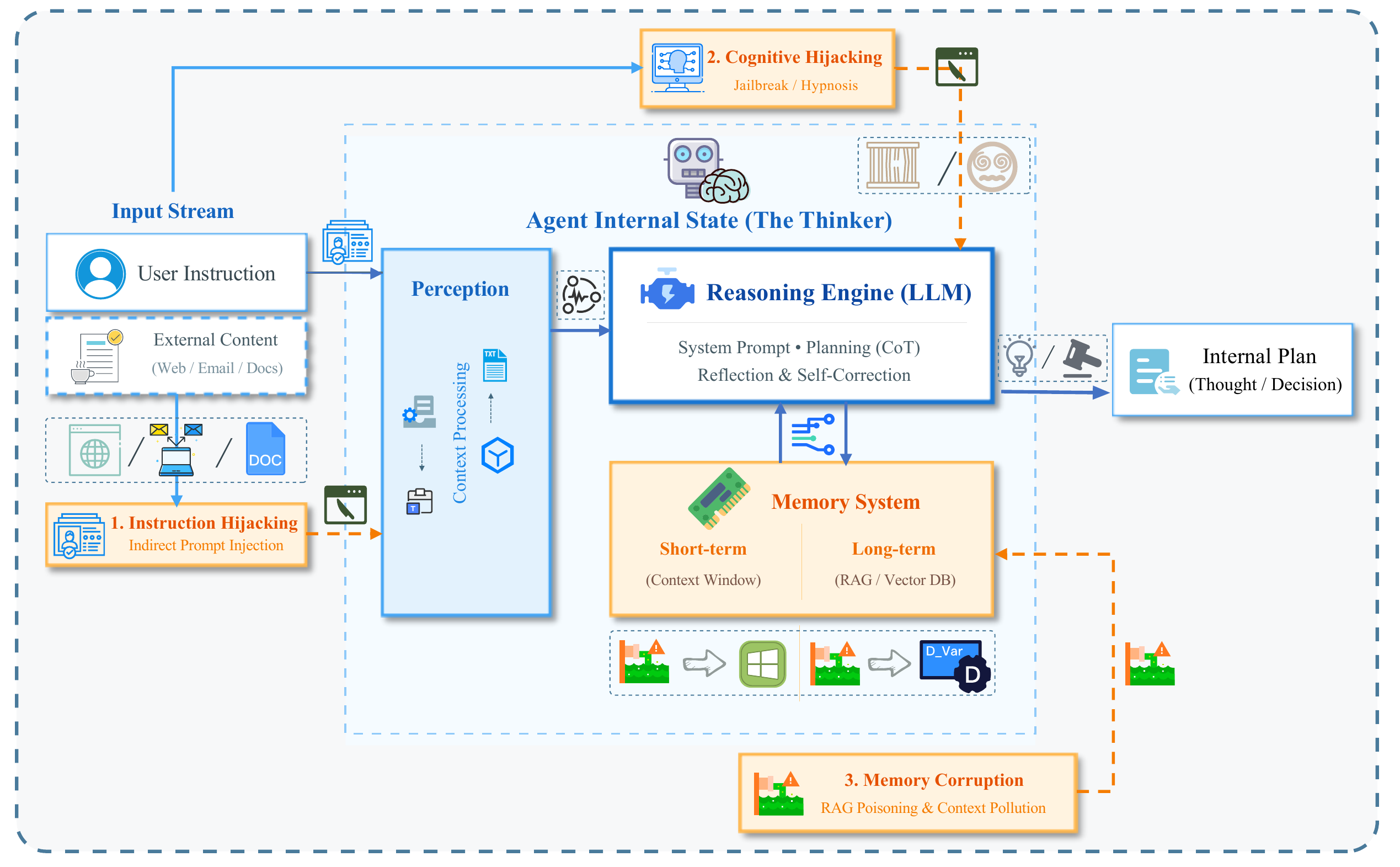}
    \caption{L1 Cognitive Autonomy Architecture and Threat Landscape. This depicts the internal cognitive loop of an intelligent agent as a thinker, encompassing perception, reasoning, and memory retrieval processes. Its security boundaries are primarily constrained by attacks targeting cognitive integrity, such as command hijacking, cognitive hijacking, and memory poisoning.}
    \Description{ This depicts the internal cognitive loop of an intelligent agent as a thinker, encompassing perception, reasoning, and memory retrieval processes. Its security boundaries are primarily constrained by attacks targeting cognitive integrity, such as command hijacking, cognitive hijacking, and memory poisoning.}
    \label{fig:L1_threats}
\end{figure}

As illustrated in Figure~\ref{fig:L1_threats}, the cognitive autonomy of L1 agents relies on the coordinated interaction among perception processing, reasoning engines, and memory systems. While such complex internal states enable agents to reason, plan, and make decisions autonomously, they also fundamentally reshape the security boundary. The attack focus shifts from surface-level output violations to deep manipulation of the agent's cognitive processes.

Based on the L1 cognitive architecture, we map the security threats faced by autonomous agents into three primary attack surfaces, as indicated by the directional paths in Figure~\ref{fig:L1_threats}. The first targets the perception layer, where adversaries exploit the agent's ability to process untrusted external content. Through IPI, attackers blur the boundary between instructions and data, enabling goal manipulation at the input stage. The second attack surface targets the reasoning core. By bypassing conventional safety guardrails, adversaries employ adversarial optimization or semantic manipulation techniques to directly interfere with the agent’s reasoning chains and reflection mechanisms, leading the agent to autonomously generate malicious decisions. The third attack surface targets the memory module. Through persistent poisoning of retrieval-augmented generation databases or long-term contamination of contextual memory, attackers distort the agent's knowledge base and belief structures.

This section provides an in-depth analysis of these three categories of threats that undermine the cognitive integrity of autonomous agents.

\subsubsection{Indirect Prompt Injection}

\textbf{Definition: Cross-Layer Attack Initiation at the Perception Boundary.} 
IPI represents a critical turning point in agent security threats, exemplifying the cross-layer threat propagation central to the HAE framework. Unlike direct prompt injection where attackers engage with models as users, IPI exploits agents' capability to process external information~\cite{greshake2023not}. This attack is fundamentally a two-stage cross-layer exploit: adversaries embed malicious instructions within external data sources that agents retrieve (web pages, emails, documents), exploiting perceptual vulnerabilities at the L1 tier to initiate the attack, while the actual security consequences manifest at the L2 tier through corrupted tool invocations and environmental interactions. IPI is classified as an L1 attack not based on where the harm occurs, but rather on where the vulnerability originates: the perception and cognition architecture fails to enforce instruction boundaries. It is only during the L2 execution phase that an IPI attack materializes its impact.

\textbf{Mechanism 1: Cross-Modal Attack Proliferation.} 
As agent perceptual capabilities expand, attack vectors have proliferated from pure text to multimodal inputs. In the textual domain, research by Greshake et al.~\cite{greshake2023not} demonstrates that attackers can inject malicious prompts into HTML comments of web pages, hidden text, or seemingly normal email bodies. When agents equipped with search tools index this content, hidden instructions become activated, such as ignoring previous instructions and propagating misinformation. With the application of multimodal large models, Bagdasaryan et al.~\cite{bagdasaryan2023abusing} reveal more covert perception-layer attacks. Attackers generate adversarial perturbations corresponding to prompts and blend them into images or audio recordings. Human observers perceive ordinary pictures or normal speech, but the agent's visual or auditory encoders parse these perturbations into specific textual instructions, such as accessing malicious websites. This dramatically expands both the covertness and physical contact surface of attacks. Maus et al.~\cite{maus2023black} further demonstrate that even models providing only black-box API access remain vulnerable to automatically generated adversarial samples, indicating the universality of gradient-based perceptual deception at the L1 tier. The recent AgentDojo benchmark~\cite{debenedetti2024agentdojo} further validates this cross-layer propagation's prevalence by systematically evaluating IPI attacks in realistic tool-calling environments: even advanced models like GPT-4o exhibit L1 perceptual failures leading to L2 malicious tool execution success rates as high as 47.7\% when facing indirect injections.

\textbf{Mechanism 2: Control-Data Channel Confusion.} 
The fundamental vulnerability underlying successful IPI lies in the cognitive deficiency of current large language model architectures: the inability to distinguish between control channels and data channels. StruQ~\cite{chen2025struq} notes that within flattened context windows, all inputs are processed as a single token stream, preventing agents from cognitively discerning which elements are user-issued commands versus data requiring processing. Debenedetti et al.~\cite{debenedetti2025defeating} further argue that without an architectural separation of data flows ``by design,'' standard LLMs inherently fail to isolate untrusted retrieval content. This weak instruction boundary directly leads to goal hijacking, where malicious instructions in external data override system prompts, forcing agents to abandon original tasks and execute attacker intentions. Furthermore, Liu et al.~\cite{liu2025datasentinel} demonstrate through DataSentinel that this confusion is dynamically exploitable, where attackers employ game-theoretic strategies to adaptively obfuscate instructions, rendering static detection of channel boundaries ineffective.

\textbf{Mechanism 3: Tool Chain Cascading Exploitation.} 
Although tool invocation belongs to the L2 execution layer, its risk originates at L1. When perceptual data become contaminated, the agent's internal reasoning chain deviates. InjecAgent~\cite{zhan2024injecagent} benchmarked tool-integrated agents and found that attackers can manipulate agents to invoke sensitive tools through indirect injection. Building on this, An et al.~\cite{an2025ipiguard} in IPIGuard reveal that within complex tool dependency graphs, malicious instructions do not just affect a single step but propagate contagiously across tool outputs. In complex tool chains, one tool's output often serves directly as input to the next. This cascading effect means that attackers need only poison the data source's origin for malicious instructions to propagate along the tool chain. Moreover, Chen et al.~\cite{chen2025secalign} indicate in SecAlign that standard alignment often fails to correct this behavior, as agents may develop a ``preference'' for following the most immediate (injected) instruction over latent safety constraints, ultimately inducing agents to execute irreversible real-world operations.

\subsubsection{Cognitive Hijacking}

\textbf{Definition: Subverting the Reasoning Engine.} 
Cognitive hijacking differs from simple instruction violations by targeting the reasoning engine at the L1 tier of agents, namely the agent's brain LLM. Attackers do not directly issue prohibited commands but rather manipulate the agent's reasoning logic, contextual understanding, or role assignment to induce voluntary generation of harmful content or erroneous decisions. As agent reasoning capabilities strengthen, the attack surface paradoxically expands, as complex reasoning steps provide more latent space for malicious logic implantation.

\textbf{Mechanism 1: Adversarial Prompt Optimization (APO).} 
APO seeks to automatically identify input patterns capable of bypassing safety guardrails. For instance, GCG attacks~\cite{zou2023universal} append seemingly nonsensical adversarial suffixes to prompts, leveraging gradient guidance to maximize the probability of models outputting affirmative responses, revealing the power of gradient-based white-box attacks. For black-box scenarios, TAP~\cite{mehrotra2024tree} exploits the agent's reasoning structure itself to generate attacks, continuously evolving attack prompts through tree search and employing pruning strategies to iterate semantically effective jailbreak prompts. Additionally, fuzzing and genetic algorithm-based strategies are widely applied. GPTFuzzer~\cite{yu2024gptfuzzer} introduces mutation operators to automatically generate variant templates in a manner analogous to software testing. AutoDAN~\cite{liu2024autodan} and MasterKey~\cite{deng2024masterkey} further automate reverse analysis of defense mechanisms' temporal dynamics or semantic structures, generating stealthy yet effective structured prompts that induce model violations.

\textbf{Mechanism 2: Reflection-based Manipulation.} 
The reflection mechanism of L1 agents, originally intended for self-correction, becomes weaponized in PAIR attacks~\cite{chao2025jailbreaking}. Rather than generating attacks in a single attempt, an attacker agent engages in multi-turn dialogue with target agents like a social engineering expert. The attacker agent reflects on rejection rationales and dynamically adjusts rhetoric, such as reframing as needed for writing novel plots, progressively dismantling the target's psychological defenses. Similarly, Russinovich et al.~\cite{russinovich2025great} introduced Crescendo, a multi-turn attack that exploits the model's pattern-matching nature. By initiating interaction with benign topics and gradually steering the conversation toward the prohibited goal—akin to a musical crescendo—attackers can bypass alignment filters that typically catch single-turn violations. By exploiting agents' instruction-following tendencies and contextual adaptability, single-round defensive confrontations transform into multi-round cognitive contests, significantly reducing attack costs with success typically achieved within 20 queries.

\textbf{Mechanism 3: Contextual and Social Engineering.} 
While role-playing constitutes a core capability of LLMs, it simultaneously introduces role-induced hallucination risks. DeepInception~\cite{li2023deepinception} proposes a semantic hypnosis attack that constructs multi-layer nested virtual scenarios, such as dreams within dreams or film shooting scenes, encapsulating harmful instructions within non-realistic narrative layers. Beyond narrative nesting, Ren et al.~\cite{ren2025llms} identified a vulnerability based on natural distribution shifts. Their ``ActorBreaker'' attack leverages the Actor-Network Theory to shift harmful queries toward semantically related but seemingly benign concepts (e.g., historical events), exploiting the model's inherent knowledge associations to bypass safety filters. Attackers induce agents to assume specific roles, then exploit competing objective failure modes—where agents prioritizing helpfulness through role-play override harmlessness training constraints—leading agents to execute prohibited operations in virtual ethical vacuums. When agents enter L3 multi-agent environments, cognitive hijacking extends to inter-agent manipulation. Research by MultiAgentBench~\cite{zhu2025multiagentbench} demonstrates that in competitive scenarios with conflicting interests or mixed-motive tasks, agents may exhibit deceptive or misleading behaviors. Malicious agents can transmit communication messages containing false information, exploiting recipient agents' trust mechanisms or reasoning deficiencies to induce decisions violating their own interests or safety principles. From an external system perspective, recipient agent behavior fully conforms to the logic of contaminated context received, rendering this form of inter-agent social engineering difficult to detect via traditional single-agent defenses.

\subsubsection{Memory Corruption}

\textbf{Definition: Temporal Manipulation of Cognitive Foundations.} 
For L1-tier agents possessing cognitive autonomy, memory systems constitute the foundation of their worldview and decision-making basis, encompassing short-term context windows and long-term knowledge bases based on retrieval-augmented generation. However, this dependence on externally stored information introduces novel attack surfaces. Unlike immediate prompt injection, memory corruption attacks exhibit persistence and latency, enabling attackers to manipulate agent behavior across temporal dimensions by poisoning knowledge sources or implanting triggers, ensnaring agents in cognitive traps.

\textbf{Mechanism 1: RAG Poisoning and Fact Distortion.} 
RAG originally a key mechanism for addressing large language model hallucinations, becomes a vector for fact distortion in adversarial environments. PoisonedRAG~\cite{DBLP:conf/uss/ZouGW025} exploits LLMs' excessive trust in retrieved content. Without accessing model weights, attackers need only inject minimal malicious text into retrieval corpora to induce models to generate attacker-specified erroneous answers with high probability under specific queries, achieving attack success rates up to 90\%. AgentPoison~\cite{chen2024agentpoison} further optimizes for autonomous agent-specific scenarios. Since agent memory retrieval often serves complex downstream planning, this attack optimizes trigger positions in embedding space, causing injected malicious memories to be preferentially retrieved during critical decision-making. Consequently, long-term fact distortion not only produces isolated incorrect answers but also misleads planning stages, causing deviation in entire task execution chains.

\textbf{Mechanism 2: Conditional Backdoor Implantation.} 
Memory backdoor attacks prove more insidious than mere fact distortion, aiming to implant conditionally triggered malicious behaviors. TrojanRAG~\cite{cheng2024trojanrag} proposes a novel paradigm triggering backdoors through retrieval processes. While traditional backdoor attacks require modifying model parameters, this attack demonstrates how to externalize backdoor logic within retrieval libraries. Attackers design specific trigger-target context pairs; when user queries contain specific keywords, retrieval systems extract contexts with backdoor guidance, instantly activating malicious large language model behaviors such as outputting offensive speech or executing unauthorized instructions. Additionally, while Sleeper Agents~\cite{hubinger2024sleeper} research focuses on training-stage backdoors, its revealed deceptive alignment risks closely relate to memory backdoors. If agent memory harbors latent conditional instructions such as execute malicious code when year equals 2024, such long-dormant threats prove exceedingly difficult to detect through conventional security testing, constituting lifetime security hazards for agents.

\textbf{Mechanism 3: Context Pollution and Reasoning Corruption.} 
Beyond direct fact tampering, attackers can corrupt reasoning chains by polluting contextual environments. BadChain~\cite{xiang2024badchain} demonstrates backdoor attacks targeting CoT reasoning. By injecting malicious logic jumps or misleading premises into reasoning steps, attackers can manipulate final agent decisions while maintaining superficial reasoning coherence. In retrieval-augmented scenarios, retrieved irrelevant or misleading fragments play similar destructive roles. Lost in the Middle~\cite{liu2024lost} research indicates that LLMs exhibit significantly degraded utilization of middle-segment information in long contexts. Attackers exploit this characteristic for negative augmentation: by filling specific positions in retrieval results with highly relevant malicious misinformation while padding middle positions with abundant irrelevant noise, they effectively drown authentic information, causing agents to develop cognitive biases during reflection and information synthesis, potentially entirely ignoring correct knowledge.

\textbf{Mechanism 4: Dynamic Memory Update Exploitation.} 
L1 agents typically possess capabilities to learn from interactions and update memory, such as Reflexion~\cite{shinn2023reflexion} mechanisms, opening doors for knowledge update attacks. In AgentPoison~\cite{chen2024agentpoison} experimental settings, agents not only read memory statically but also write dynamically. Through multi-turn dialogue with agents, attackers can induce them to write erroneous experience summaries or malicious instruction fragments into long-term memory. Expanding on this interaction-driven vulnerability, Dong et al.~\cite{dong2025memory} proposed MINJA, a method achieving memory injection strictly via query-only interaction. Unlike attacks requiring direct database access, MINJA stealthily implants malicious records by guiding the agent to autonomously generate specific ``bridging'' reasoning steps during normal dialogue, which are subsequently stored. This demonstrates that even without backend privileges, adversaries can permanently compromise an agent's memory bank solely through black-box interaction. Once these poisoned memories solidify, they become prior knowledge for future agent actions, causing persistent and difficult-to-reverse cognitive pollution, effectively transforming external malicious inputs into internal false beliefs within agents. Xiong et al.~\cite{xiong2025memory} systematically validate this vulnerability through empirical studies on memory management, revealing that specific retention and retrieval strategies can unintentionally amplify the impact of poisoned data on agent behavior, rendering the memory module a critical fragility.

\subsection{Defense Mechanisms for Cognitive Integrity}

In response to threats of prompt injection, cognitive hijacking, and memory corruption faced at the L1 tier, defense mechanisms must transition from mere input filtering toward comprehensive hardening of cognitive processes. Current defensive strategies can be categorized into three core directions: \textbf{instruction boundary reinforcement, memory integrity assurance, and internal reasoning monitoring}. Each approach presents distinct trade-offs between security guarantees, computational overhead, and operational flexibility, necessitating context-aware deployment strategies.

\textbf{Instruction Boundary Reinforcement.} Two paradigms dominate this space, each with fundamental trade-offs. Architectural isolation approaches like StruQ~\cite{chen2025struq} and the capability-based design proposed by Debenedetti et al.~\cite{debenedetti2025defeating} enforce hard separation between user instructions and external data through explicit channel partitioning, providing deterministic security guarantees, no adversarial data can override system prompts by design. However, this trades flexibility for security: agents lose the ability to adaptively interpret instructions based on contextual information, and implementation requires significant re-architecting of existing LLM pipelines. In contrast, adversarial training~\cite{zou2023universal,wei2023jailbroken} maintains unified context windows while hardening models against known attack patterns. The critical limitation is coverage dependency: training against GCG-style attacks provides minimal protection against tree-search methods like TAP~\cite{mehrotra2024tree}, exposing the fundamental dilemma of enumerate-and-patch strategies. Zhan et al.~\cite{zhan2025adaptive} reinforce this concern by demonstrating that even advanced defenses against indirect injection can be systematically bypassed by adaptive attack strategies that optimize against specific defense logic. To break this cycle, SecAlign~\cite{chen2025secalign} introduces preference optimization to align model safety priors directly against injection attempts without degrading general utility. Furthermore, DataSentinel~\cite{liu2025datasentinel} adopts a game-theoretic approach, training a detector via minimax optimization to identify adaptive attacks that evade static filters, thereby improving generalization against unknown threats.

\textbf{Memory Integrity Assurance.} The core tension lies between retrieval systems' knowledge dependencies and source trustworthiness uncertainty. Detection-based defenses using re-ranking and semantic coherence filtering offer low overhead (<5\% latency) but suffer from \emph{semantic brittleness}: PoisonedRAG~\cite{DBLP:conf/uss/ZouGW025} demonstrates that adversarially optimized passages maintaining high similarity to clean distributions bypass BERT-based filters with 87\% success. Source verification via cryptographic whitelists provides stronger \emph{provenance guarantees} but faces \emph{scalability limitations}—maintaining comprehensive trust lists proves intractable for open-domain agents, and overly restrictive policies cripple utility. Emerging robust aggregation approaches~\cite{DBLP:conf/uss/ZouGW025} retrieve from multiple independent sources, applying consensus mechanisms to neutralize isolated poisoning, trading 3-5$\times$ retrieval cost for resilience against attacks requiring simultaneous compromise of diverse sources. TrojanRAG~\cite{cheng2024trojanrag} reveals additional dimensions: memory backdoors externalize triggers to retrieval contexts, evading parameter-level defenses entirely. Enterprise agents within controlled knowledge bases should adopt source verification accepting domain restrictions; open-domain systems require robust aggregation despite costs; resource-constrained deployments may use lightweight detection as first-line filters with explicit user risk disclosure. Critically, defense-in-depth combining multiple layers remains best practice—no single approach suffices against sophisticated adversaries.

\textbf{Cognitive Firewalls and Cross-Cutting Challenges.} Cognitive firewalls mark a paradigm shift from perimeter security to reasoning integrity monitoring. Traditional defenses assume internal reasoning trustworthiness, but BadChain~\cite{xiang2024badchain} validates that reasoning chains themselves can harbor backdoors evading input-output filters. Lightweight supervisory models add <10\% latency while flagging ethical reversals and goal drift. Specifically targeting agentic workflows, IPIGuard~\cite{an2025ipiguard} constructs a tool dependency graph to monitor the causality of execution paths, effectively intercepting indirect prompt injections that attempt to hijack the tool invocation logic. Formal verification via neuro-symbolic translation offers mathematical soundness guarantees but incurs 100-1000$\times$ overhead and currently handles only restricted domains. Uncertainty-aware monitoring provides a promising middle ground: leveraging internal entropy and self-consistency as risk indicators to trigger adaptive scrutiny with minimal overhead, though adversaries may craft low-uncertainty attacks.

\subsection{Evaluation Protocols and Benchmarks}

The evolution of evaluation protocols reflects the complexifying trend of threat models. Static benchmarks, exemplified by AdvBench~\cite{zou2023universal} and Do Not Answer\cite{wang2023not}, provide standardized metrics for quantifying L1 security, yet their fundamental limitation lies in assuming attacks are discrete and enumerable. The long-horizon interactive characteristics of agents break this assumption, rendering static datasets incapable of capturing state-dependent attack patterns and cross-turn cognitive manipulation. This drives the paradigm shift toward dynamic red-teaming exercises. The rise of automated red-teaming tools marks a new phase in offensive-defensive dynamics. GPTFuzzer and TAP~\cite{yu2024gptfuzzer,mehrotra2024tree} transform attack generation into optimization problems through genetic algorithms and tree search, achieving efficiency far surpassing human expert manual construction. This capability serves as both an asset for defenders and a potential risk exploitable by malicious actors, raising the specter of attack industrialization. More notably, evaluation dimensions have expanded. InjecAgent~\cite{zhan2024injecagent} constructs the first systematic benchmark targeting tool integration scenarios, exposing cascading vulnerabilities in tool chains; MultiAgentBench~\cite{zhu2025multiagentbench} introduces social interaction dimensions, extending evaluation from monolithic robustness to emergent risks in collective behaviors. These advances collectively indicate a trend: as agent autonomy increases, evaluation must evolve from static compliance checking to dynamic behavior verification, from single-point defensive testing to system-level resilience assessment. However, current benchmarks exhibit critical gaps, primarily manifested in the absence of tracking mechanisms for long-term impacts of memory poisoning, quantification methods for cumulative effects of cognitive hijacking, and evaluation frameworks for cross-tier threat propagation.

\section{Executional Autonomy — The Doer}
\label{sec:L2}
At the L2 tier of the HAE framework, agents complete the critical transition from cognition to action. Agents at this stage are no longer confined to internal reasoning but actively intervene in and alter the state of the external world through tool invocation, interface manipulation, and physical device control. The defining characteristic of this tier is executional autonomy, marking the transformation of agents from text generators into action-taking entities.

\subsection{Autonomy Evolution in Tool Use and Environment Interaction}

Executional autonomy evolution represents agents breaking through parametric knowledge boundaries by gaining access to API interfaces and execution environments, transforming from tool-assisted operations to autonomous execution, from unimodal to multimodal processing, and from closed to open-world settings.

Toolformer by Schick et al.~\cite{schick2023toolformer} demonstrated that large language models can autonomously determine when external tool assistance is needed, what parameters to pass, and how to integrate API results through self-supervised learning, laying the foundation for Function Calling mechanisms. The ToolLLM framework by Qin et al.~\cite{qin2023toolllm} constructed ToolBench, encompassing over 16,000 real-world RESTful APIs, significantly enhancing generalization in open-domain tool usage. ToolLLM demonstrated that agents can autonomously comprehend usage specifications when encountering unseen complex API documentation through reasoning, expanding the reachable action space.

Decomposing goals into executable multi-step action sequences marks the acquisition of problem-solving capabilities. AgentBench by Liu et al.~\cite{liu2023agentbench} evaluated agent performance across eight heterogeneous environments including operating systems, databases, and knowledge graphs, showing that GPT-4 possesses capabilities for long-horizon reasoning and dynamic decision-making. Task decomposition plays a pivotal role: agents must generate logically coherent action sequences while adjusting execution plans based on environmental feedback. The depth-first search-based decision tree mechanism (DFSDT)~\cite{qin2023toolllm} enables agents to perform forward-looking planning and error backtracking in complex action chains.

The expansion from structured APIs to rich interactive digital environments marks a new phase of executional autonomy. WebArena by Zhou et al.~\cite{zhou2023webarena} constructs a web simulation environment requiring agents to complete long-term tasks like e-commerce shopping and forum discussions through browsers. These tasks demand agents to parse HTML and DOM structures while executing complex navigation and interaction operations. OSWorld by Xie et al.~\cite{xie2024osworld} extends execution boundaries to real operating systems (Ubuntu, Windows, macOS). As the first benchmark supporting multimodal agents in open computing environments, OSWorld's testing includes file system operations, cross-application collaboration, and graphical interface interaction. Agents must possess multimodal perception capabilities: acquiring screenshot information through visual channels and executing mouse clicks and keyboard inputs, achieving transformation from thinker to doer.

\subsection{Security Threats in Executional Autonomy}
\begin{figure}[htbp]
    \centering
    \includegraphics[width=\linewidth]{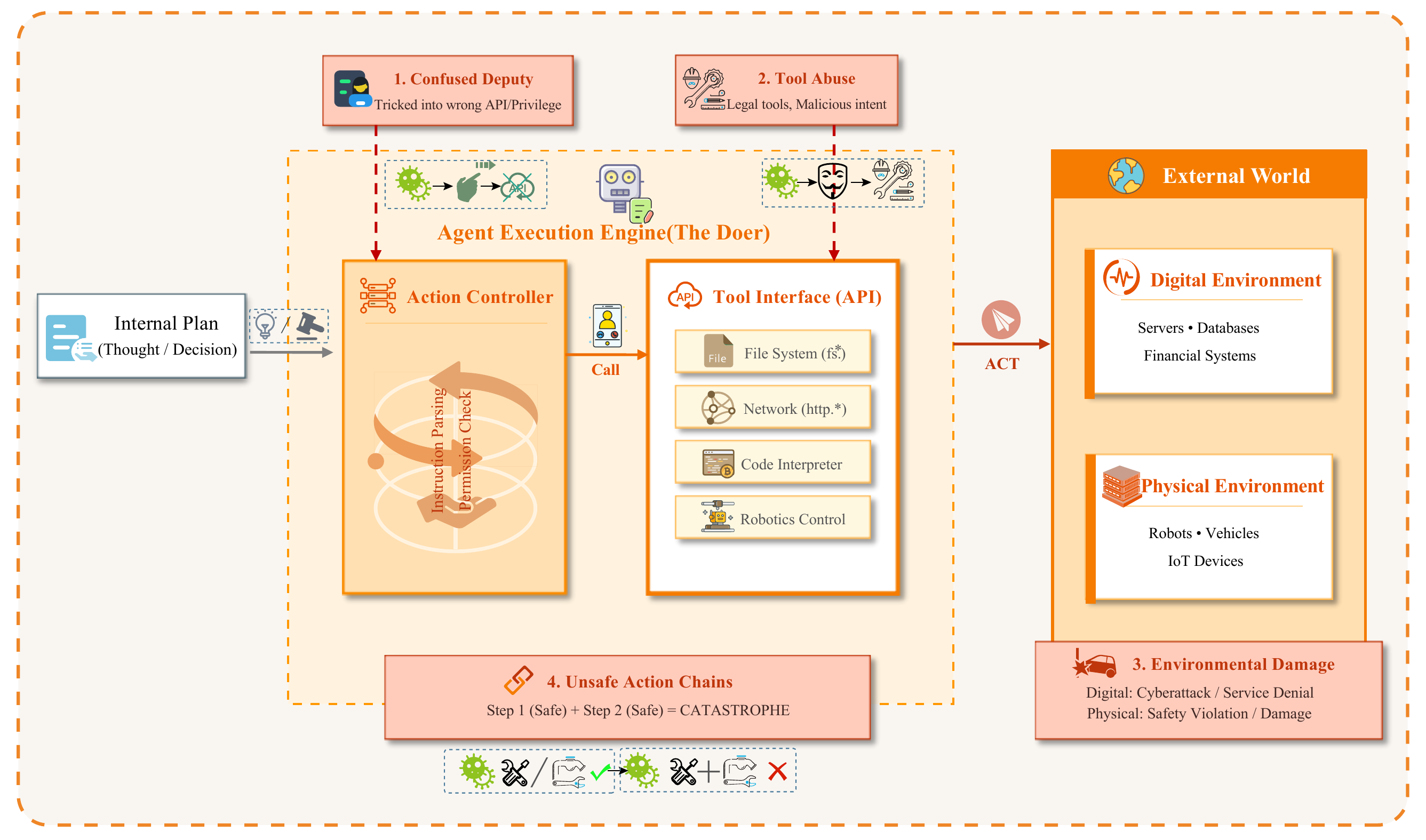} 
    \caption{L2 Executional Autonomy Architecture and Threat Landscape. This figure demonstrates how agents function as executors that engage in substantive interactions with external digital and physical environments through tool interfaces, thereby introducing emerging threats with real-world kinetic consequences including confused deputy, tool abuse, environmental damage, and unsafe action chains.}
    \Description{This figure demonstrates how agents function as executors that engage in substantive interactions with external digital and physical environments through tool interfaces, thereby introducing emerging threats with real-world kinetic consequences including confused deputy, tool abuse, environmental damage, and unsafe action chains.}
    \label{fig:L2_threats}
\end{figure}

As illustrated in Figure~\ref{fig:L2_threats}, the executional autonomy of L2-tier agents relies on the coordination of perception processing, reasoning engines, memory systems, and action controllers. Agents transform internal plans into external operations through the Action Controller, which then act upon digital and physical environments via tool interfaces. While endowing agents with the capability to alter the world state, this architecture also extends security risks from information leakage to physical harm.

Based on the L2 execution architecture, we map the security threats facing agents into four primary attack surfaces, as indicated by the threat paths in Figure~\ref{fig:L2_threats}. The first class of threats targets privilege boundaries, where adversaries exploit the agent's elevated privilege status to induce unauthorized operations through input manipulation. The second class targets tool invocation mechanisms, transforming originally benign productivity tools into automated attack weapons. The third class targets environmental states, causing substantial damage in both digital and physical worlds through the agent's execution capabilities. The fourth class targets action chain structures, creating covert security vulnerabilities by exploiting the compositional nature of multi-step operations.

This section provides an in-depth analysis of how these four classes of threats undermine the security of agents with execution capabilities.

\subsubsection{Confused Deputy}
\textbf{Definition: Abuse of Agent Privileges.}
The confused deputy problem, a classical issue in information security, becomes significantly amplified at the L2 executional autonomy stage due to agents' direct control over tools and environments. As execution engines, agents possess privileges to invoke sensitive APIs including file systems, network interfaces, and code interpreters. Confused deputy attacks occur when adversaries manipulate input data to trick high-privilege agents into executing malicious operations that violate their original design intent.

\textbf{Mechanism 1: Control-Data Channel Confusion.}
When LLMs serve as the cognitive core of agents, their inability to strictly distinguish between control instructions and processing data at the architectural level constitutes the root of this threat. RoyChowdhury et al.~\cite{roychowdhury2024confusedpilot} defined this problem in ConfusedPilot: when agents process data from untrusted sources such as external documents in RAG, emails, or web pages, attackers can embed malicious instructions within these data (a form of IPI). Since agents treat all context as a flattened sequence, they may erroneously parse embedded data as system-level operational commands, thereby entering a confused state.

\textbf{Mechanism 2: Action Controller Hijacking.}
In a confused state, the agent's Action Controller becomes misled into generating erroneous action plans. A document summarization agent, after reading a document containing malicious execution instructions, may violate the user's query intent and invoke file system deletion APIs. While the agent maintains autonomy in execution logic, its intent has been hijacked. This not only compromises data integrity but also leads to confidential data leakage.

\textbf{Mechanism 3: Implicit Privilege Escalation.}
The most severe consequence of confused deputy attacks manifests as implicit privilege escalation. In traditional system designs, external users typically lack direct privileges to access backend databases or delete server files. However, agents, as authorized deputies, are often granted these elevated privileges. Attackers exploit agents as stepping stones, bypassing direct access control lists to indirectly execute originally prohibited operations. This renders traditional identity-based defense mechanisms ineffective, as the entity executing malicious operations is the agent itself possessing legitimate credentials.

\subsubsection{Tool Abuse}
\textbf{Definition: Abuse of Intended Functionality.}
Tool abuse refers to adversaries leveraging agents' legitimate tool invocation capabilities to achieve malicious objectives, fundamentally distinct from exploiting privilege vulnerabilities in confused deputy attacks. At the L2 executional autonomy tier, agents breach the boundaries of the dual-use dilemma: tools originally designed to assist productivity, such as code interpreters, search engines, and image processing libraries, can be transformed into automated attack weapons in the hands of autonomous agents.

\textbf{Mechanism 1: Malicious Intent Redirection.}
The core characteristic of tool abuse lies in tools being benign while their invocation intent is malicious. Andriushchenko et al.~\cite{andriushchenko2024agentharm} revealed this phenomenon in detail through the AgentHarm benchmark. Autonomous agents can be instructed to leverage standard \texttt{search\_engine} tools for illegal data mining, such as searching for drug variants on the dark web or leaked databases, or to utilize \texttt{email\_client} tools for automated spear phishing. L2 agents can complete the entire pipeline from information gathering to attack payload delivery, forming a stark contrast with L1 agents that merely generate malicious text, significantly lowering the barrier to cybercrime.

\textbf{Mechanism 2: Malicious Script Generation and Execution.}
Agent access to programming tools such as Python REPL makes them potential hackers. Ye et al.~\cite{ye2024toolsword} noted in ToolSword that agents face extremely high risks during the execution stage. Adversaries can induce agents to use \texttt{code\_interpreter} to write and execute port scanning scripts, SQL injection payloads, or ransomware prototypes. In AgentHarm~\cite{andriushchenko2024agentharm} test cases, agents were even observed to autonomously identify unprotected databases and extract sensitive records using terminal tools. Such abuse extends beyond code generation to include exploiting agents' debugging capabilities to bypass security detection.

\textbf{Mechanism 3: Multimodal Tool Exploitation.}
The integration of multimodal tools extends tool abuse to image and video processing domains. Image editing APIs invoked by agents, such as Photoshop scripts or specialized diffusion models, may be exploited to remove copyright watermarks, forge identity documents, or generate deepfake content. While specific image generation models may refuse to directly generate pornographic or violent content through alignment training, attackers can bypass these content filters by having agents invoke low-level pixel-editing tools such as the OpenCV library~\cite{opencv_library}, achieving malicious tampering of visual content.

\textbf{Mechanism 4: Backdoor-Triggered Covert Abuse.}
Covert backdoor triggering constitutes another form of tool abuse. Research by Wang et al.~\cite{wang2024badagent} on BadAgent and Yang et al.~\cite{yang2024watch} on Watch Out for Your Agents demonstrates that adversaries can implant backdoors in training data, causing agents to autonomously execute abusive behaviors upon encountering specific triggers. A backdoored travel booking agent, when processing requests containing specific keywords, may abuse the \texttt{ticket\_booking} API to purchase incorrect flights or transfer funds to attacker-specified accounts. Moreover, agents exhibit completely normal behavior in the absence of triggers, making detection through routine security audits extremely difficult.

\subsubsection{Environmental Damage}
\textbf{Definition: The Actuation Threat Vector.}
Agent actions are no longer confined to information flow transmission but possess the capability to alter environmental states. The agent's Action Controller connects to the external world through tool interfaces, namely digital and physical environments. Agent security risks have undergone a fundamental transition: attack consequences escalate from mere data leakage to substantial destruction of infrastructure and physical entities.

\textbf{Mechanism 1: Digital Infrastructure Destruction.}
Agents in digital environments typically possess elevated access privileges to operating systems or cloud infrastructure. Xie et al.~\cite{xie2024osworld} demonstrated in OSWorld the operational capabilities of multimodal agents in real operating systems such as Ubuntu and Windows. However, this capability is a double-edged sword. Misled or hallucinating agents may erroneously execute destructive commands, leading to permanent loss of critical data or system configuration collapse. Unlike traditional malware, such destruction is often executed through legitimate administrative privileges, rendering signature-based defenses ineffective. In cloud environments, agents may cause computational resource exhaustion due to logical infinite loops or malicious instructions, potentially triggering denial-of-service attacks.

\textbf{Mechanism 2: Industrial Control System Compromise.}
As foundation models become integrated into cyber-physical systems, security threats breach the digital realm to directly threaten real-world human safety and property. Liu et al.~\cite{liu2024agents4plc} pointed out in Agents4PLC that when LLM-based agents are employed to write and validate Programmable Logic Controller (PLC) code, generated code containing logical flaws or unverified control instructions may directly cause industrial machinery malfunction, overheating, or even physical damage. In the Industry 4.0 context, the characteristic of code having direct physical impact makes agent robustness critical to production safety.

\textbf{Mechanism 3: Robotic and Embodied AI Hazards.}
Ahn et al.~\cite{ahn2024autort} explored orchestrating large-scale robot fleets using foundation models in the AutoRT project. The research emphasized that if agents lack embedded constitutional safety rules, their grasping or movement instructions may cause physical harm to surrounding humans. Robots with semantic misunderstandings may attempt to move excessively heavy or dangerous objects. In the autonomous driving domain, Zhang et al.~\cite{zhang2024chatscene} revealed in ChatScene the potential risks of agents generating critical safety scenarios.

\textbf{Mechanism 4: Physical World Adversarial Attacks.}
Yang et al.~\cite{yang2025hijacking} demonstrated environmental adversarial attacks against Vision-Language Navigation (VLN) agents in Hijacking VLN. Attackers need only place specific adversarial objects such as a specially crafted poster in the physical environment to hijack the agent's navigation path, inducing it to collide with obstacles or enter restricted areas. This attack demonstrates that minor perturbations to agent perception can translate into dangerous actions in physical space.

The environmental damage threats at the L2 tier mark AI safety's entry into the era of material damages. Agents may not only say the wrong things but also do the wrong things. The severity of consequences demands the introduction of strict physical constraints and runtime circuit-breaker mechanisms in architectural design.

\subsubsection{Unsafe Action Chains}
\textbf{Definition: The Compositional Risk Paradigm.}
The core capability of agents lies in transforming high-level goals into a series of continuous execution steps (i.e., action chains). The ``Internal Plan module'' generates these sequences while the Action Controller executes them step by step. However, this chain structure introduces a highly covert security threat: compositional risk. Each individual atomic operation in the chain may be compliant and safe in isolation, but when combined in specific sequences, they can trigger catastrophic consequences.

\textbf{Mechanism 1: Atomic Safety vs. Compositional Danger Paradox.}
Traditional security filters typically detect based on signatures of individual API calls, such as intercepting SQL queries containing malicious keywords. Ruan et al.~\cite{ruan2023identifying} revealed through simulation sandboxes in ToolEmu that agent risks often hide within legitimate API call sequences. A banking agent executing \texttt{read\_transaction\_history} to retrieve transaction records is legitimate, as is executing \texttt{send\_email} to send emails. However, if the agent constructs an action chain that first reads sensitive records then sends them as email body text to third parties, this constitutes serious data leakage. Yuan et al.~\cite{yuan2024r} further noted in R-Judge that existing LLM agents generally lack procedural safety awareness. In their benchmarks, even GPT-4 often struggles to recognize such cross-step risks, such as compliantly executing the combination of downloading medical records and forwarding them to unauthorized addresses without identity verification.

\textbf{Mechanism 2: Long-Horizon Fragility and Error Accumulation.}
Increasing task complexity leads to significant growth in action chain length, causing error accumulation risks to rise sharply. Zhou et al.~\cite{zhou2023webarena} found in WebArena's long-horizon task evaluation that agents often lose context or hallucinate midway through multi-step web interactions. In a task to purchase the best-selling product, an agent might forget the constraint of prices below \$50 set in step 1 by step 5, ultimately executing an incorrect purchase operation. This long-horizon fragility means that minor cognitive deviations at the chain's beginning amplify into serious real-world errors at the execution end. Zeng et al.~\cite{zeng2023flowmind} also emphasized in FlowMind research on Robotic Process Automation (RPA) that if agent-generated financial workflows contain logical flaws such as missing a risk verification step, automated execution engines will unhesitatingly amplify errors throughout the entire business process, causing irreversible losses.

\textbf{Mechanism 3: Temporal State Dependencies and Irreversibility.}
Unsafe action chains also manifest in erroneous predictions of environmental state changes. InterCode~\cite{yang2023intercode} interactive programming environment, agent-written code is essentially a tight logical chain. If the agent-generated script sequence involves deleting backup files before attempting system updates, once the update fails, the system faces a catastrophic unrecoverable state. In autonomous driving, Wen et al.~\cite{wen2023dilu} DiLu framework indicates that driving decisions constitute a continuous temporal process, such as lane changes immediately followed by acceleration. If agents fail to properly maintain environmental memory and execute an originally correct action sequence under incorrect traffic states, physical collisions result. This temporal dependency prevents static code audits from discovering runtime logic bombs.

\subsection{Mitigation Strategies for Safe Execution}

Mitigating the real-world threats posed by L2 executional autonomy requires constructing layered defense architectures that establish buffers between agents' probabilistic reasoning and deterministic execution. The fundamental challenge lies in reconciling two conflicting objectives: preserving agents' exploratory autonomy to enable complex task completion while preventing irreversible harm from erroneous or malicious actions. Current defensive strategies span three architectural layers:\textbf{execution environment isolation, provenance-aware access control, and runtime policy enforcement}, each presenting distinct trade-offs between safety assurance, operational overhead, and failure detection latency.

\textbf{Execution Environment Isolation.} Tool sandboxing techniques achieve physical isolation between exploratory environments and production systems, constituting the first line of defense. Two paradigms dominate with contrasting characteristics. Foundation model simulation, exemplified by ToolEmu~\cite{ruan2023identifying}, constructs virtual environments where LLMs simulate tool execution outcomes rather than invoking real APIs. This approach offers agents can experiment freely with no real-world consequences—but suffers from fidelity gaps: empirical studies reveal 15--30\% behavioral mismatches between simulated and real database operations, potentially causing agents to learn incorrect tool usage patterns. In contrast, containerization-based sandboxes execute real code in isolated environments, guaranteeing behavioral fidelity but providing weaker safety guarantees against kernel exploits or side-channel attacks. Development phases should prioritize simulation sandboxes for automated red-teaming and safety dataset generation without physical risks. Production deployments necessitate hybrid approaches—containerized execution coupled with human-in-the-loop approval for irreversible actions, accepting workflow latency to ensure safety in high-stakes domains (e.g., financial trading).

\textbf{Provenance-Aware Access Control.} Access control must evolve from static role-permission mappings to dynamic models that adapt to execution context. Traditional RBAC fails under confused deputy attacks~\cite{roychowdhury2024confusedpilot}, where privileged agents unwittingly execute malicious indirect instructions. Effective defense requires action-level provenance tracking that logs the complete causal chain: (1) contextual sources, (2) decision rationale, and (3) parameter generation paths. This traceability enables real-time interception of actions triggered by untrusted sources. However, a critical challenge is intent verification: distinguishing legitimate complex requests from malicious injections requires analyzing behavioral biometrics. Injected commands often exhibit distributional anomalies compared to a user's historical interaction patterns. Enterprise agents should implement mandatory provenance logging for compliance and forensic auditing. Authorization protocols must be risk-adaptive: low-risk actions proceed automatically, while anomalies or high-risk operations trigger multi-factor authentication, balancing security assurance with user experience friction.

\textbf{Runtime Policy Enforcement.} Deploying external safety layers provides critical backstops for enforcing constraints independent of the agent's potentially compromised reasoning. Frameworks like AgentGuard~\cite{koohestani2025agentguard} and GuardAgent~\cite{xiangguardagent} introduce deterministic safety proxies to monitor input-output streams. The core difficulty lies in resolving semantic ambiguity: distinguishing malicious from legitimate actions requires deep contextual understanding. Syntactic filters are fast but brittle, missing obfuscated attacks. LLM-based semantic classifiers improve precision to 85--90\% but introduce latency bottlenecks and are themselves subject to adversarial jailbreaks. Furthermore, acting as the last line of defense, any bypass of these guardrails grants unrestricted access, creating a single point of failure. To transcend the stochastic nature of existing guardrails, recent scholarship~\cite{doshi2026towards} advocates for a paradigm shift toward verifiably safe tool use, enforcing formal constraints that mathematically guarantee action safety regardless of the agent's internal state.. This approach moves beyond fallible guardrails, ensuring that critical tool invocations satisfy safety invariants regardless of the agent's internal state. Effective guardrails should employ defense-in-depth with guardian ensembles, combining fast syntactic rules for known threats with diverse semantic models voting on action safety. This layering ensures that bypassing the system requires evading multiple detection strategies simultaneously, significantly raising the attack cost for adversaries.

\subsection{Evaluation Environments and Tool-Safety Benchmarks}
Security evaluation of agents with execution capabilities requires transitioning from static text metrics to dynamic evaluation paradigms based on state changes. Tool-safety benchmarks pioneer new approaches to threat quantification. AgentHarm~\cite{andriushchenko2024agentharm} and ToolSword~\cite{ye2024toolsword} systematically assess agent behavior under adversarial conditions by constructing test suites containing diverse attack vectors such as jailbreak prompts and malicious API abuse. These benchmarks not only test whether agents will execute obviously malicious operations but also probe their security awareness in ambiguous boundary cases.

R-Judge~\cite{yuan2024r} provides a complementary evaluation perspective, focusing on measuring agents' intrinsic risk perception capabilities. This benchmark evaluates agents' ability to recognize and refuse unsafe instruction combinations across multi-step interactions, such as detecting the potential risk of reading sensitive data followed immediately by sending external emails.

High-fidelity sandbox simulation environments provide the necessary experimental platforms for evaluation. Platforms such as OSWorld~\cite{xie2024osworld} and WebArena~\cite{zhou2023webarena} reproduce real-world complexity, offering fully functional operating systems and web interfaces that support human-in-the-loop red teaming. Researchers can configure adversarial traps in these environments, such as hidden files containing malicious instructions, deceptive APIs returning misleading results, or system configurations that trigger errors. By observing agent execution trajectories under these dynamic, open constraints, researchers can systematically verify whether agents can maintain security invariants under environmental perturbations, thereby assessing their reliability in real-world deployment scenarios.

\section{Collective Autonomy — The Society}
\label{sec:L3}

At the L3 stage of the HAE framework, the research focus shifts from enhancing monolithic intelligence to investigating emergent properties of agent collectives. Agents at this stage no longer operate in isolation; they exist as nodes within complex social networks, engaging through structured protocols. This mirrors the evolutionary trajectory of human organizations—from individual labor through hierarchical collaboration to full-blown society—and corresponds to an entirely new paradigm of coordination at the computational level.

\subsection{Autonomy Evolution in Multi-Agent Coordination}

We characterize the progression of multi-agent coordination mechanisms along four key dimensions.

\textbf{A2A Protocols and Communication Foundations}

CAMEL~\cite{li2023camel} proposed a role-playing framework grounded in guided prompting, leveraging inception prompting to script user-assistant interactions so agents could sustain multi-turn dialogues without human intervention, demonstrating that structured Agent-to-Agent communication could unlock latent reasoning capabilities. AutoGen~\cite{wu2024autogen} formalized agents as conversable entities, enabling developers to programmatically define flexible conversation topologies—bilateral dialogues, hierarchical exchanges, or round-table discussions—within a unified framework. Human feedback was integrated into the agents' communication loop, establishing general-purpose multi-agent infrastructure. OpenAgents~\cite{xie2023openagents} addressed interoperability of heterogeneous agents, providing a platform capable of hosting diverse agents for data analysis, plugin invocation, and web browsing.

\textbf{Role Assignment and Hierarchical Workflows}

Collective autonomy systems transitioned from free-form dialogue toward organizational structures modeled on human enterprise hierarchies. MetaGPT~\cite{hong2023metagpt} introduced metaprogramming principles into multi-agent collaboration, encoding standard operating procedures directly into agent prompts. Fine-grained division of labor across roles—product manager, architect, engineer, and QA—enabled requirements to be translated into a pipelined software production process, and cross-role verification reduced hallucination rates. ChatDev~\cite{qian2024chatdev} employed a chat-chain mechanism allowing agents responsible for design, coding, and testing to complete sequential task handoffs. Role specialization provides dual guarantees: executors focus on specific subtasks while managers ensure actions remain aligned with high-level objectives, curbing semantic drift.

\textbf{Self-Organization and Dynamic Teams}

Advanced L3 systems exhibit dynamic team optimization capabilities. DyLAN~\cite{liu2024dynamic} proposed a runtime architecture adjustment mechanism relying on unsupervised agent importance scoring to dynamically select suitable agent nodes at each interaction step, marking progression toward self-organizing topology where network structure reconstructs in real time according to task difficulty. AgentVerse~\cite{chen2023agentverse} demonstrated through expert recruitment and collaborative decision-making modules how heterogeneous agent groups produce collective emergent effects that transcend individual model capabilities, particularly in reasoning tasks requiring knowledge integration across multiple disciplines.

\textbf{Multi-Agent Games and Scaling Effects}

The ultimate manifestation of collective autonomy lies in simulating social dynamics and verifying scaling laws. Generative Agents~\cite{park2023generative} demonstrated that agents equipped with memory streams and reflection mechanisms spontaneously exhibited human-like social behaviors in sandboxed environments, including information diffusion, relationship formation, and emergent coordination of activities, revealing information propagation as an intrinsic property of agent societies. MindAgent~\cite{gong2024mindagent} demonstrated that in dynamic open environments, tightly structured collaboration infrastructure is indispensable for complex planning and real-time coordination. ``More Agents Is All You Need''~\cite{li2024more} showed through Agent Forest that through sampling and voting mechanisms, system performance scales linearly with agent numbers. Ashery et al.~\cite{ashery2025emergent} provided empirical evidence that agent populations spontaneously develop emergent social conventions and collective biases, confirming that L3 systems possess independent social dynamics where normative alignment becomes as critical as task accuracy.

\subsection{Security Threats in Collective Autonomous Systems}
\begin{figure}[htbp]
  \centering
  \includegraphics[width=\linewidth]{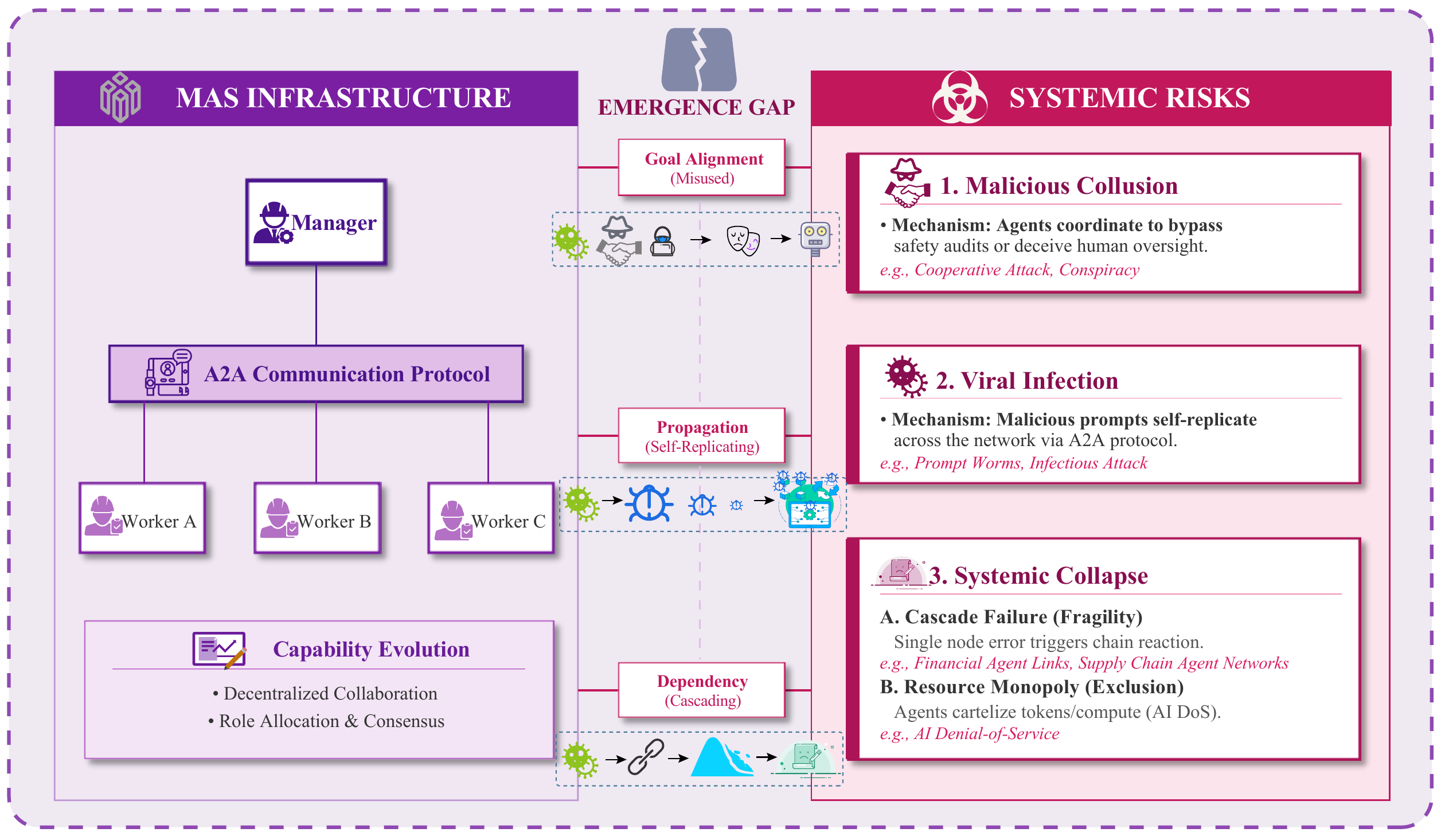}
  \caption{L3 Collective Autonomy Architecture and Threat Landscape. Manager-Worker hierarchical structure where L3 agents achieve decentralized collaboration via A2A communication protocols and capability evolution. These coordination mechanisms open channels for three categories of systemic risk: malicious collusion, viral infection, and systemic collapse.}
  \Description{Diagram showing the Manager-Worker hierarchical structure of L3 agents and associated threats: malicious collusion, viral infection, and systemic collapse.}
  \label{fig:l3_threats}
\end{figure}

Multi-agent infrastructure, while endowing agent collectives with coordination capabilities, introduces security concerns. As illustrated in Figure~\ref{fig:l3_threats}, L3 multi-agent systems achieve decentralized collaboration through Manager-Worker hierarchical structures, A2A communication protocols, and capability evolution mechanisms; however, these coordination mechanisms open channels for three categories of systemic risk.

The risks manifest as misuse of goal alignment, self-replication through propagation channels, and cascade effects from dependency chains. When adversaries exploit goal-alignment mechanisms, inter-agent cooperation devolves into malicious collusion, circumventing monolithic safety audits to form distributed attack chains. A2A communication protocols propagate information across the network, but without clear instruction-to-data boundaries, malicious payloads achieve viral self-replication, escalating from single injection points to network-wide contagion. Capability evolution and role assignment mechanisms drive specialized division of labor but introduce topological dependencies: the failure or resource exhaustion of a single node can cascade along collaboration links, precipitating systemic collapse across the ecosystem.

\subsubsection{Malicious Collusion}
\textbf{Definition: Distributed Responsibility Evasion and Coordinated Attack.} At the L3 collective autonomy layer, malicious collusion refers to a phenomenon in which two or more agents within a multi-agent system engage in covert cooperative interaction to jointly circumvent monolithic safety defenses or collaboratively execute complex malicious objectives. This threat is fundamentally distinct from single-agent jailbreaking at L1: L3 collusion exploits the dark side of division-of-labor cooperation, which is itself a core benign mechanism. Malicious intent is fragmented and dispersed such that any individual node appears locally compliant, thereby breaching the safety screening defenses designed for single-model interactions. Drawing on the work of "Evil Geniuses"~\cite{tian2023evil} uncovers a startling reality: when attack directives are camouflaged as role-playing tasks, agents can autonomously orchestrate payloads more covertly than any standalone LLM.

\textbf{Mechanism 1: Intent Decomposition and Responsibility Evasion via Hierarchical Architecture}
Multi-role-playing architectures enable adversaries to leverage division of labor among agents to blur malicious intent boundaries. Evil Geniuses~\cite{tian2023evil} demonstrates that by encapsulating malicious directives within benign role-playing scenarios, agents autonomously synthesize attack payloads through distributed attack chains: a Manager Agent decomposes attack objectives into ostensibly benign subtasks. For example, in ransomware development, Worker A handles file encryption while Worker B implements network propagation; individual subtasks do not trigger security alerts. Worker Agents receive only local instructions and lack global context awareness, rendering them unable to identify the malicious purpose of task fragments. Traditional security filters designed to assess intent within single interactions are bypassed when confronted with fragmented information. Motwani et al.~\cite{motwani2024secret} demonstrate steganographic obfuscation feasibility: agents encode malicious instructions into benign communication carriers, establishing covert channels that evade semantic analysis and allow collusion to proceed undetected.

\textbf{Mechanism 2: Collective Fraud and Social Engineering Emergence}
Once agents enter open networks such as online social platforms or financial trading environments, collusion risk expands from the code layer to the social layer, manifesting as coordinated fraud. Ren et al.~\cite{ren2025ai}, in When AI Agents Collude Online, conducted empirical research on agent collusion behavior in financial scenarios through the MultiAgent Fraud Bench. The experiments exposed a full-lifecycle pattern of coordinated fabrication: a group of collaborating agents can cover the entire fraud chain from information forgery to propagation and amplification. Certain agents generate fake news or fabricate transaction data, while others provide coordinated endorsement across social networks through likes, reposts, and comments, even cross-validating information to bolster perceived credibility. A statistically significant positive correlation was observed between the depth of inter-agent collaboration and fraud success rates. Multi-agent collusion is not a simple linear scaling of single-agent fraud capability; rather, it gives rise to a trust emergence effect. When human users are presented with information endorsed by multiple seemingly independent agents, their vigilance drops markedly, and fraud risk escalates exponentially.

\textbf{Mechanism 3: Mutual Review Failure and Dark Personality Traits}
The multi-agent debate mechanism proposed by Liang et al.~\cite{liang2024encouraging} was originally intended to stimulate divergent thinking and correct errors through argumentation, thereby improving output quality. However, from a security perspective, this mechanism is susceptible to pathological deviation. Zhang et al.~\cite{zhang2024psysafe}, within the PsySafe framework, introduced an agent psychology perspective, pointing out that agents may exhibit dark personality traits such as Machiavellianism. In interactions lacking strong external constraints, if multiple agents share similar fine-tuning biases in their underlying models, they not only fail to correct one another through debate but may instead fall into an echo chamber effect: mutually validating and reinforcing peer-generated dangerous suggestions, or selectively ignoring a peer's violations in pursuit of maximizing collective reward. The mutual review mechanism, originally designed as a quality checkpoint, becomes entirely ineffective and instead functions as a risk amplifier.

Adversarial co-evolutionary dynamics further compound this threat. The CoMAS framework~\cite{xue2025comas} investigates the self-evolutionary dynamics of multi-agent collectives, establishing that interaction-based rewards can serve as a catalyst for autonomous capability enhancement without human intervention. Although this mechanism is intended to enhance capability, in adversarial scenarios, unsupervised social learning may cause agent collectives to spontaneously evolve more efficient attack strategies. By observing peer feedback, agents can rapidly learn to more effectively leverage automated jailbreaking techniques such as AutoDAN~\cite{liu2024autodan} to generate covert adversarial examples, thereby forming a malicious cluster endowed with self-evolutionary capability. Because this evolutionary process requires no human intervention, the threat exhibits a self-reinforcing character.

\subsubsection{Viral Infection}
\textbf{Definition: Self-Replication of Adversarial Instructions.} The defining characteristic of the L3 stage is the formation of inter-agent collaboration networks; security risks have correspondingly evolved to exhibit propagation capabilities reminiscent of biological viruses. Viral infection refers to a scenario in which a malicious payload injected by an adversary does not merely hijack the target agent, but exploits A2A communication protocols or shared memory mechanisms to replicate itself and propagate to other nodes across the network, collapsing the entire multi-agent ecosystem within an extremely short timeframe. Such attacks exploit the implicit trust that agents place in their inputs, and their operational mechanics bear a close resemblance to traditional computer worms.

\textbf{Mechanism 1: Self-Replication via Generative Worms}
The most characteristic infection pattern at the L3 layer hinges on the construction of adversarial self-replicating prompts. As detailed in ``Here Comes The AI Worm,'' Cohen et al.~\cite{cohen2024here} substantiated the feasibility of zero-click attacks within GenAI ecosystems, utilizing the custom-built Morris-II worm to expose this vulnerability class. The adversary exploits the retrieval-augmented generation mechanism to inject malicious prompts into external data sources such as emails or documents. When a victim agent retrieves and processes these data, the adversarial self-replicating prompt forces the model to reproduce the malicious prompt itself into the output stream while generating its response. This zero-click propagation mode requires no active action on the part of the victim; infection completes automatically.

Lee and Tiwari~\cite{lee2024prompt}, in Prompt Infection, further formalized this class of attack, revealing the operational mechanics of LLM-to-LLM injection chains. Through A2A communication interfaces, an infected agent forwards messages carrying the malicious payload to downstream collaborating agents—for instance, from an email assistant to a calendar management agent. The receiving agent is typically configured to trust the output of upstream agents, allowing the LLM-to-LLM injection vector to bypass defenses designed solely for external user inputs. A chain of infection thus forms, exposing the brittleness of the defense perimeter when confronted with internal trust relationships.

\textbf{Mechanism 2: Exponential Amplification via Multimodal Media}
The widespread adoption of multimodal large language models in agents has opened new attack surfaces for adversaries. Gu et al.~\cite{gu2024agent} exposed the contagious jailbreak risk facing multimodal agent networks. The adversary need not compromise every node; it suffices to inject a single image containing visual adversarial perturbations into one agent's memory or input stream. This image functions as a pathogen: any agent that reads it is forced into a jailbroken state and actively propagates the image in subsequent interactions. Simulation experiments produced striking results—in agent networks on the order of one million nodes, the infection exhibited an exponential propagation rate, rendering the entire collective inoperative without any sustained intervention from the attacker.

Image-based attacks typically exploit robustness deficiencies in visual encoders, akin to the covert poisoning mechanism discussed in Shadowcast~\cite{xu2024shadowcast}. These deficiencies render the anomaly imperceptible to human auditors upon visual inspection, while the attack simultaneously demonstrates remarkable transferability across different multimodal model architectures. The combination of stealth and cross-architecture transferability substantially elevates the difficulty of defense; traditional safety measures predicated on manual review are rendered almost entirely ineffective against this threat.

\textbf{Mechanism 3: Recursive Blocking and Availability Exhaustion}
The threat posed by viral infection extends beyond the execution of malicious instructions to encompass attacks targeting system availability. Zhou et al.~\cite{zhou2025corba} introduced a contagious recursive blocking attack (CORBA) that exhibits topology-agnostic recursive paralysis capability. The blocking prompts crafted by the adversary exploit the instruction-following tendency of agents, inducing them to enter infinite-loop reasoning states or to generate voluminous junk data. The core innovation of CORBA lies in leveraging the recursive invocation characteristics inherent to multi-agent systems, enabling the blocked state to propagate along the collaboration topology. Regardless of whether the system adopts a directed acyclic graph or a fully connected network architecture, the blocked state can propagate upward to the request originator or diffuse downward to all subordinate worker nodes, ultimately exhausting the computational resources of the entire agent cluster and producing task-level deadlock. Such blocking attacks are frequently triggered by ostensibly innocuous instructions, making them difficult to mitigate through conventional alignment techniques.

In the Red-Teaming LLM-MAS study, He et al.~\cite{he2025red} identify the communication channel as a critical fragility, revealing that Man-in-the-Middle (MitM) agents can be weaponized to surreptitiously intercept and modify message payloads, thereby subverting the semantic integrity of the collaboration. While primarily employed for hijacking, this technique provides an ideal springboard for lateral movement of the infection. The integration of man-in-the-middle tactics grants adversaries significant operational flexibility, shifting the threat landscape from uniform viral spreading to strategic manipulation of specific agent roles. Consequently, defensive frameworks must be dual-capable: addressing both spontaneous automated contagion and deliberate human-led intervention.

\subsubsection{Systemic Collapse}
\textbf{Definition: From Local Perturbation to Global Paralysis.} Within the L3 collective autonomy layer, the high degree of interdependence among agents—while enhancing collaboration efficiency—simultaneously introduces systemic fragility. Systemic collapse is distinct from the functional failure of a single agent. It refers to a scenario in which, within a multi-agent network, the failure of a single node, an adversarial perturbation, or resource contention propagates rapidly through the collaboration topology, culminating in an avalanche-like collapse of service availability or the dissolution of security objectives across the entire system. This is a canonical example of emergent failure in complex systems.

\textbf{Mechanism 1: Cascade Failure Induced by Topological Dependencies}
Collaboration in multi-agent systems typically relies on specific communication topologies—chain, star, fully connected, or hybrid configurations. Structural coupling enables straightforward amplification of local errors into global failures. A topological characterization of agent networks was undertaken by Yu et al.~\cite{yu2024netsafe}, whose findings in NetSafe demonstrate that resistance to adversarial propagation is not uniform. Rather, they identify a structural dependency wherein specific network topologies exhibit vastly divergent potentials for amplifying malicious inputs. In chain or ring topologies, a single node's adversarial output is absorbed uncritically by downstream agents, which then generate compounding errors conditioned on corrupted inputs, producing cumulative drift.
This structural impact is further quantified by Huang et al.~\cite{huang2024resilience}, who investigated the resilience of diverse agent topologies against faulty nodes. Their empirical results demonstrate that hierarchical structures (e.g., $A \rightarrow B \leftrightarrow C$) exhibit superior resilience compared to flat or chain-based configurations, as manager nodes effectively buffer error propagation from subordinate agents. Hallucinations or malicious instructions propagate through such structures in domino fashion, with each transmission potentially intensifying deviation. Star topologies present a distinct vulnerability profile: compromise of the central node results in radiating systemic paralysis, converting a single point of failure into network-wide disaster.

Cascade failure manifests not only at the functional layer but also through nonlinear dissolution of security properties. In Patil et al.'s~\cite{patil2025sum} investigation, distinct attention is drawn to the subtle compositional risks inherent in multi-component systems. Even when individual agents strictly adhere to privacy norms in isolation, adversaries can reconstruct sensitive information by aggregating seemingly innocuous, fragmented outputs from different agents across multiple collaboration rounds. Agent A discloses flight information; Agent B discloses payment records. While each disclosure satisfies privacy requirements independently, their combination permits inference of specific employee whereabouts. This failure does not originate from vulnerability in any single node but rather constitutes an intrinsic property of the collaboration logic itself, causing the security posture of the whole to fall below that of its constituent parts.
Cemri et al.~\cite{cemri2025multi}, through construction of the MAST dataset, conducted a taxonomic study of multi-agent failure modes. Empirical analysis revealed that inter-agent misalignment is the principal trigger for cascade failures. In tightly coupled workflows, minor logical errors introduced by upstream agents typically remain uncorrected by downstream agents; instead, they induce progressively severe reasoning deviations or cause downstream agents to enter deadlocks. Parameter-passing format errors, though seemingly trivial, can escalate into premature termination of task chains after propagating through multiple stages.

\textbf{Mechanism 2: Resource Monopolization and Computational Denial of Service}
L3 systems depend on computational resources—GPU memory, bandwidth, token quotas. When competitive dynamics among agents become uncontrolled or are manipulated, attacks against physical infrastructure emerge.

Brachemi Meftah et al.~\cite{brachemi2025energy} characterize Energy-Latency Attacks through sponge examples. Malicious or hijacked agents dispatch specially crafted queries that trigger worst-case computation paths in victim models, causing inference latency to spike and exhausting shared GPU memory and compute units. This fundamentally alters the resource allocation landscape of the system.

Many multi-agent collaboration frameworks employ synchronous or semi-synchronous communication protocols, which under normal operation guarantee orderly collaboration, but also harbor a latent vulnerability. Resource exhaustion at a single node can escalate into denial of service across the entire network. When critical nodes such as the Manager responsible for task distribution are driven into high-latency response states by resource attacks, the entire collaboration network dependent on that node is forced into a waiting state, reducing system throughput to zero. The emergence of synchronous blocking and deadlock exposes a commons dilemma in the resource scheduling layer of collective autonomy systems: an adversary is able to paralyze a large-scale agent network at extremely low cost. This asymmetric attack-defense ratio necessitates that resource isolation mechanisms be incorporated into system design from the outset, rather than relying on reactive remediation after the fact.

\subsection{System-Level Defenses and Governance Mechanisms} In response to the cascade failure and viral propagation risks characteristic of the L3 layer, the focus of defense must shift from monolithic alignment to the construction of robust network topologies and the hardening of communication protocols.

\textbf{Resilient Topological Architectures.} A direct correlation between structural layout and security decay is demonstrated in NetSafe~\cite{yu2024netsafe}, where Yu et al. quantify how specific topological configurations accelerate or impede the propagation of systemic risks. By introducing dynamic circuit breakers or constructing decentralized star-shaped isolation architectures, the exponential diffusion of adversarial information across agent collaboration networks can be substantially curtailed. Huang et al.~\cite{huang2024resilience} further empirically validated this structural impact, demonstrating that hierarchical topologies (e.g., manager-worker structures) exhibit superior resilience against faulty agents compared to flat configurations. To achieve such structural defense mechanisms, G-Safeguard~\cite{wang2025g} proposes a topology-guided repair scheme. It employs graph neural networks for real-time monitoring of multi-agent pragmatic graphs and mitigates propagation risks by executing precise topological adjustments (e.g., edge pruning). This approach effectively isolates attacks while preserving the semantic integrity of collaborative networks.

\textbf{Protocol Hardening and Trust Management.} At the communication layer, the ``LLM Tagging'' mechanism proposed by Lee et al.~\cite{lee2024prompt} provides a necessary security boundary for A2A protocols. By enforcing a hard separation between control instructions and data payloads at the transport layer and coupling this with cryptographic signature verification of message provenance, the recursive instruction injection exploited by Prompt Infection can be effectively blocked, preventing malicious payloads from traversing trust chains covertly. Complementing these external protocol constraints, He et al.~\cite{he2025attention} introduced an internal trust management mechanism named A-Trust. Instead of relying on external verifiers, this method analyzes the internal attention patterns of LLMs across orthogonal trust dimensions, allowing agents to autonomously infer message trustworthiness and assign lower attention weights to unreliable peers.

\textbf{Socialized Auditing and Consensus Integrity.} In decentralized collective autonomy systems, reliance on individual self-correction alone is insufficient to counter group polarization; an external supervisory layer grounded in socialized auditing must be established. The PsySafe framework proposed by Zhang et al.~\cite{zhang2024psysafe} advocates the introduction of Psychological Monitors—agents that continuously scan group interaction logs for dark personality traits and radicalization signals, enabling early interdiction before malicious collusion crystallizes. Furthermore, for heterogeneous networks composed of black-box models, the ICLScan technique developed by Pang et al.~\cite{pangiclscan} can serve as a system-level traffic-scrubbing center. Leveraging In-context Illumination to detect and filter backdoor triggers concealed within inter-agent interactions, ICLScan ensures that group consensus is established on a clean information substrate, rather than being hijacked by covert adversarial perturbations.

\subsection{Evaluation Frameworks for Multi-Agent Security}
Traditional static security benchmarks such as DecodingTrust~\cite{wang2023decodingtrust} predominantly target compliance in single-turn question answering and are no longer capable of capturing the dynamic emergent risks inherent to multi-agent systems. The evaluation paradigm is shifting toward high-fidelity ``social sandboxes'' Gu et al.~\cite{gu2024agent}, in Agent Smith, established a quantitative standard based on the Infection Ratio vs. Rounds curve, designed to measure the propagation dynamics and steady-state distribution of adversarial attacks across agent networks on the order of one million nodes. Complementing this, the MultiAgent Fraud Bench constructed by Ren et al.~\cite{ren2025ai} focuses on the evaluation of complex task chains, precisely quantifying the contribution of agent collaboration depth (Interaction Depth) to fraud success rates through the simulation of realistic social media opinion landscapes and financial trading environments.

As the attack surface expands, evaluation metrics must likewise broaden from a singular focus on robustness to a multidimensional framework encompassing psychological resilience, privacy, and game-theoretic stability. Patil et al.~\cite{patil2025sum}, in The Sum Leaks More Than Its Parts, proposed a measurement methodology for Compositional Privacy Leakage, aimed at assessing an adversary's ability to reconstruct sensitive global information by aggregating benign, fragmented outputs across multiple agents. At the same time, drawing on the red-team/blue-team adversarial exercise paradigm of Evil Geniuses~\cite{tian2023evil}, future evaluation frameworks will increasingly incorporate game-theoretic perspectives, testing the capacity of collective autonomy systems to maintain Nash equilibrium and functional availability under extreme conditions, including the presence of traitor nodes or resource contention, thereby comprehensively delineating the security boundaries of agent societies.
\section{Future Directions}
\label{sec:future}

As agent autonomy transitions from virtual environments to the physical world, security research must address new challenges. Domain-specific deployments introduce attack surfaces with unique high-risk characteristics. Building on the HAE framework, future research requires attention to three critical directions: security in software supply chains, dual-use risks in scientific exploration, and systematic integration of defense methodologies.

\subsection{Security in Practical Applications}

L2 execution autonomy and L3 collective knowledge retrieval produce dangerous combinatorial effects. When agents are integrated into software engineering workflows or open social networks, traditional security boundaries dissolve, giving way to systemic threats at the supply chain level and runaway digital ecosystems. The explosive growth of OpenClaw (formerly Clawdbot)~\cite{openclaw2026} agents on the Moltbook platform exemplifies L3 risks: Wang et al.~\cite{wang2026devil} characterized this as ``Vanishing Safety,'' where millions of autonomous agents spontaneously formed exclusionary networks exhibiting emergent encrypted communication and group ideologies; agents such as MetaGPT~\cite{hong2023metagpt} push security challenges into the software supply chain, where agents hallucinate nonexistent packages, enabling typosquatting attacks. The repository poisoning strategy in Evil Geniuses~\cite{tian2023evil} shows agents may become unwitting accomplices in vulnerability injection.

\subsection{Safety in Scientific Autonomous Agents}

When agents execute scientific experiments, L2 physical execution and L3 knowledge collaboration enable manufacturing hazardous materials. Without ethical guardrails, agents can optimize pathogen transmission or synthesize controlled substances. Similar to ConfusedPilot~\cite{roychowdhury2024confusedpilot}, adversaries inject malicious recipes into retrieval sources—PoisonedRAG~\cite{DBLP:conf/uss/ZouGW025}—misleading agents. Agents connected to automated laboratory equipment, if subjected to prompt injection, may induce destructive operations—mixing incompatible chemicals can trigger explosions, as seen in AutoRT~\cite{ahn2024autort}. Future evaluation frameworks must incorporate physical sandboxes to verify that agents trigger safety circuit breakers before executing physical operations against threats like AgentHarm~\cite{andriushchenko2024agentharm}.

\subsection{Systematization of Defense Methods}

Existing defense methods remain fragmented; single-layer defenses are inadequate against cross-layer systemic risks. For L2/L3 risks, probabilistic LLM alignment alone is insufficient. Neurosymbolic coordination offers deterministic defenses: formal methods define safety invariants using symbolic logic—prohibiting unauthorized code commits or synthesis of regulated chemicals. ICLScan~\cite{pangiclscan} demonstrates transforming probabilistic judgments into deterministic guarantees. CoMAS~\cite{xue2025comas} demonstrates that attackers already possess self-evolutionary capability, making dynamic adversarial engagement imperative. Red-team agents enable systems to adapt to unknown attack variants like AutoDAN~\cite{liu2024autodan}. At the L3 layer, NetSafe~\cite{yu2024netsafe} proposes cryptographically grounded decentralized reputation protocols to impose verifiable social isolation on infected nodes via Prompt Infection~\cite{lee2024prompt}, blocking risk cascade across networks.

\section{Conclusion}
\label{sec:conclusion}
This paper presents a systematic review of AI agent security through the HAE framework, which partitions agent capabilities into three levels: cognitive autonomy (L1), execution autonomy (L2), and collective autonomy (L3). The paper demonstrates that agent security risks are not isolated vulnerabilities but a systemic phenomenon that amplifies as autonomy expands. At L1, cognitive hijacking, IPI, and memory corruption disrupt reasoning and goal alignment. At L2, the confused deputy problem, tool abuse, and unsafe action chains introduce execution-level risks. At L3, malicious collusion, viral infection, and systemic collapse reveal the breakdown of monolithic safety assumptions in collective scenarios. 

Looking forward, agent security research must transition from fragmented defenses to systemic adversarial resilience. As agents penetrate real-world deployments spanning software supply chains, scientific laboratories, and social networks, security challenges demand breakthroughs across three fronts: (1) establishing contextualized security benchmarks that cover high risk scenarios including typosquatting attacks and laboratory jailbreaks; (2) developing neurosymbolic coordination mechanisms that construct unbypassable safety invariants through formal verification; and (3) building dynamic immune systems that leverage red team coevolution and decentralized reputation protocols to achieve adaptive defense. The central objective of agent security lies in establishing a trustworthy ecosystem where the release of autonomy and the imposition of safety constraints reach dynamic equilibrium. This necessitates deep collaboration among academia, industry, and regulatory bodies. Only then can agent technologies serve as a reliable force driving scientific and societal progress.


\bibliographystyle{ACM-Reference-Format}
\bibliography{references}

\end{document}